\documentclass[prd,aps,twocolumn,nobalancelastpage,nofootinbib,preprintnumbers,superscriptaddress,longbibliography,floatfix]{revtex4-2}
\usepackage[utf8]{inputenc}
\usepackage[colorlinks=true,citecolor=blue,linkcolor=blue]{hyperref}
\usepackage[normalem]{ulem}
\usepackage{amsmath,amssymb, mathrsfs,xfrac}
\usepackage{epsfig}
\usepackage{graphicx}
\usepackage{slashed}
\usepackage{multirow}
\usepackage{placeins}
\usepackage[dvipsnames]{xcolor}
\usepackage{natbib}
\usepackage{epstopdf}
\usepackage{soul}
\usepackage{tikz}
\usepackage[capitalise, english]{cleveref}
\usepackage{siunitx}
\usepackage{xspace}
\usepackage{booktabs}
\usetikzlibrary{trees}
\usetikzlibrary{decorations.pathmorphing}
\usetikzlibrary{decorations.markings}
\usepackage{lettrine}
\input Zallman.fd

\LettrineTextFont{\itshape}
\newcommand\myshade{80}
\colorlet{mylinkcolor}{ForestGreen}
\colorlet{mycitecolor}{Aquamarine}
\colorlet{myurlcolor}{violet}

\newcommand{\vect}{\boldsymbol}

\hypersetup{
  linkcolor  = mylinkcolor!\myshade!black,
  citecolor  = mycitecolor!\myshade!black,
  urlcolor   = myurlcolor!\myshade!black,
  colorlinks = true
}

\definecolor{jblue}{RGB}{20,50,100}
\definecolor{npurple}{RGB} {153, 51, 204}
\definecolor{wred}{RGB}{217,0,56}
\definecolor{white}{RGB}{255,255,255}

\definecolor{korange}{RGB}{235, 80,  43}
\definecolor{korange2}{RGB}{245, 100,  63}
\definecolor{kyelloworange}{RGB}{255, 210,  110}
\definecolor{kyelloworange2}{RGB}{240, 170,  90}
\definecolor{kred}{RGB}{204,  102, 153}
\definecolor{kpurple}{RGB}{153,  61, 190}
\definecolor{kpurplelight}{RGB}{213,  161, 230}

 \definecolor{tobycolour}{rgb}{.5,.0,.5}

\DeclareSIUnit\year{yr}
\DeclareSIUnit\pc{pc}
\DeclareSIUnit\ergs{ergs}
\DeclareSIUnit\msun{\ensuremath{M_\odot}}
\allowdisplaybreaks

\newcommand{\vesc}{\ensuremath{v_{\rm esc}}}

\newcommand{\keV}{\,\mathrm{keV}}
\newcommand{\GeV}{\,\mathrm{GeV}}
\newcommand{\TeV}{\,\mathrm{TeV}}
\newcommand{\mchi}{m_\chi}

\newcommand{\lz}{LZ230616}
\newcommand{\kms}{\,\mathrm{km\,s^{-1}}}
\newcommand{\vmin}{v_{\rm min}}

\newcommand{\apjl}{ApJ Lett.}
\newcommand{\apjs}{ApJS}
\newcommand{\aap}{A\&A}

\definecolor{lime}{HTML}{A6CE39}
\DeclareRobustCommand{\orcidicon}{\hspace{-1mm}
	\begin{tikzpicture}
	\draw[lime, fill=lime] (0,0)
	circle [radius=0.16]
	node[white] {{\fontfamily{qag}\selectfont \tiny \,ID}};
	\draw[white, fill=white] (-0.0525,0.095)
	circle [radius=0.007];
	\end{tikzpicture}
	\hspace{-3mm}
}

\foreach \x in {A,..., Z}{\expandafter\xdef\csname orcid\x\endcsname{\noexpand\href{https://orcid.org/\csname orcidauthor\x\endcsname}
			{\noexpand\orcidicon}}
}

\newcommand{\mytitle}{GAIA meets LZ: a high velocity tail-tale sign for dark matter }

\begin{document}

\title{\mytitle}

\author{Anirban Das\orcidA{}}
\email{anirbandas.21@protonmail.com}
\affiliation{Theoretical Physics Division, Saha Institute of Nuclear Physics, 1/AF, Bidhannagar, Kolkata 700064, India}
\affiliation{Homi Bhabha National Institute, Training School Complex, Anushaktinagar, Mumbai 400094, India}

\author{Subhabrata Majumdar \orcidB{}}
\email{subha@tifr.res.in}
\affiliation{Tata Institute of Fundamental Research, 1 Homi Bhabha Road, Colaba, Mumbai 400005, India}

\author{Manibrata Sen \orcidC{}}
\email{manibrata@iitb.ac.in}
\affiliation{Department of Physics, Indian Institute of Technology Bombay, Powai, Mumbai 400076, India}

\author{Amogh Srivastav \orcidD{}}
\email{23b1826@iitb.ac.in}
\affiliation{Department of Physics, Indian Institute of Technology Bombay, Powai, Mumbai 400076, India}

\date{\today}

\begin{abstract}
The interpretation of high-energy nuclear recoils in direct-detection experiments can depend sensitively on the high-velocity tail of the galactic dark matter (DM) distribution.
The LUX-ZEPLIN (LZ) experiment has recently reported an event, LZ230616, with nuclear recoil energy 248 $\pm$ 23 (stat) $\pm$ 23 (sys) keV, while the Gaia mission has provided an unprecedented view of the kinematics and mass distribution of the Milky Way. We bring these two probes together to investigate the inelastic DM interpretation of LZ230616, using a velocity distribution inferred from Gaia-informed galactic model. Compared to the Standard Halo Model, the resulting distribution has a lower escape speed and a significantly suppressed high-velocity tail, with a higher local density. We find that while the exothermic interpretation is largely insensitive to this change, the endothermic interpretation requires substantially larger scattering cross sections and smaller mass splittings. Furthermore, using simulation-based study of the gravitational effects of the Large Magellanic Cloud, we show that the high-velocity tail can extend the allowed endothermic parameter space towards lower DM masses and larger splittings. Our results demonstrate how combining direct searches for DM detection with precision measurements of galactic dynamics can qualitatively affect the particle-physics interpretation of high-energy nuclear recoils.

\end{abstract}
\maketitle

\noindent \textbf{\emph{Introduction}--}
The search for dark matter (DM) through its scattering off atomic nuclei provides a direct route for probing the particle nature of DM. The interpretation of a nuclear recoil, however, depends not only on the underlying microscopic interactions of DM, but also the properties of the DM halo in which the Milky Way (MW) is embedded. In particular, the velocity distribution of DM in the solar neighbourhood determines the fraction of particles capable of producing a given recoil. This becomes especially important for scattering processes that require DM particles to lie in the high velocity tail of the galactic distribution.

The recent observation of a high energy nuclear recoil event by the LUX-ZEPLIN (LZ) collaboration offers an interesting setting to explore this interplay between particle physics and galactic dynamics\,\cite{LZ:2026axp}. The event, designated LZ230616, has a reconstructed recoil energy 248 $\pm$ 23 (stat) $\pm$ 23 (sys) keV. While it is too preliminary to conclude that this is an evidence of the detection of DM, the high recoil energy makes it a useful case study for investigating whether this can arise from inelastic DM models~\cite{Nagata:2026pbj, An:2026pkc, yin2026, bandyopadhyay2026, fan2026, jeesun2026, rodd2026, Ge:2026xax, Xing:2026civ, Borah:2026ris, Ghosh:2026txe, Das:2026buc, Lian:2026hpm, Ahmed:2026kan, He:2026idw, Nguyen:2026lui, Chatterjee:2026scv, Fan:2026hzw, Qi:2026vyp, Kumar:2026lgi, Baer:2026fpy, Yuan:2026djt, Cheung:2026byg, Lee:2026jxl, Asadi:2026iot, Langhoff:2026ujr, Okada:2026eol, Ahmed:2026qjg, Du:2026lpa, Borah:2026zwf, Bisal:2026khf, Bose:2026ndd, Bose:2026szs, Wang:2026ytg,DiMauro:2026dqp, Das:2026uyy, Lee:2026wof, Dent:2026bji, Gu:2026vto,deLima:2026shq, Smirnov:2026aqk, Du:2026guj, Pospelov:2026ewn,Visinelli:2026kgt, Nomura:2026qyq, Yamashita:2026ump, Su:2026rwz, Freese:2026sga, Wu:2026nhi},  boosted dark matter~\cite{ Mahapatra:2026glu, Heikinheimo:2026kwp, Alhazmi:2026efz, Kannike:2026qyl, Liang:2026coz}, and other phenomenological models~\cite{dimauro2026,mccabe2026, Lueiza-Colipi:2026gij, Arcadi:2026kev, Barman:2026omh, Palmisano:2026kuj, Uttayarat:2026isp, DiMauro:2026ymt, Okada:2026upm, Lee:2026zbr, Chattaraj:2026fxn, He:2026hqz, Egorov:2026dpr, Bamwidhi:2026vdu, Elahi:2026vlm,Zhu:2026dag,Aghaie:2026vsu, Lee:2026xxh, Khan:2026nwp, Yang:2026wpb, Kotlarski:2026pep,Unwin:2026rdp, Brdar:2026ukx}. Crucially, this interpretation relies on the availability of sufficiently fast-moving DM particles.

Direct detection analyses conventionally assume the Standard Halo Model (SHM), in which the galactic DM population is modelled by an isotropic, truncated Maxwellian velocity distribution. This prescription has been used widely for comparing experimental results and constraining DM interactions. Nevertheless, the SHM represents an idealised description of the galactic halo, and its characteristic velocity scale, the escape speed, and the local DM density, are adapted from different astrophysical inputs. The actual phase-space distribution of DM is determined by the dynamical evolution of the MW halo under gravitational interactions, and need not correspond to the SHM. These considerations necessitate the use of observationally motivated descriptions of the DM halo, especially when analysing DM direct detection and indirect detection experimental results.

Determining the local DM density in the solar neighborhood and its velocity distribution function (VDF) has been proven to be a formidable task. The most common technique is by using the observed local stellar kinematics data. Although the stars and the DM reside within the same gravitational potential of the MW, they may have different velocity distributions. This can be used to construct the DM phase-space distribution from an observationally constrained galactic mass model through techniques such as Eddington inversion~\cite{Sayan2019}.
In recent years, the Gaia satellite has measured the positions and proper motions of more than a billion stars in the MW with high precision. This dataset has been instrumental in inferring the galactic rotation curve and mass distribution over a broad range of radii. The resulting distribution provides a physically motivated alternative to independently specified SHM parameters.

The combination of precision astrometric measurements from Gaia with direct searches for DM detection provides a unique opportunity to connect a putative DM signal to the properties of the DM halo. The consequences of this approach are particularly relevant for inelastic DM scattering. In endothermic scattering, DM particles must possess enough kinetic energy to transition to the heavier state (with mass $\mchi+\delta$), and hence prefers more DM particles in the high velocity tail of the distribution. The inferred mass-splitting $(\delta)$ and interaction strength are, therefore, sensitive to the maximum DM speed accessible in the laboratory frame. For exothermic scatterings, where the DM transitions to a lighter state (with mass $\mchi-\delta$), the release of internal energy allows for nuclear recoils, even when the incoming DM particle has lower velocities. In this case, the recoil spectrum demonstrates a characteristic peak, whose location remains comparatively stable under changes in DM velocity distributions. These properties make the LZ event useful to study how astrophysical modelling affects the interpretation of inelastic DM interpretation.

In this work, we interpret the \lz{} event using DM  velocity distributions constructed from a Gaia-informed model of the MW. We use a stellar catalogue combining Gaia DR3 astrometry with complementary spectroscopic information to constrain the galactic rotation curve, and use halo-plus-baryon mass models to determine the gravitational potential. We find that the resulting lower escape speed of DM, compared to the SHM, substantially alters the endothermic interpretation, while leaving the exothermic interpretation relatively unchanged. We further investigate the impact of the Large Magellanic Cloud (LMC), whose gravitational interaction with the MW can enhance the high-velocity tail. Including this effect further extends the endothermic parameter space towards lower DM masses and higher splittings. These findings illustrate the importance of astrophysically informed velocity distributions in interpreting high energy nuclear recoil events in direct detection experiments. Recently, Ref.\,\cite{OHare:2026nqi} also expanded on the effect of simulation-inferred VDF on the DM interpretation of the \lz{} event, emphasizing the importance of understanding the high-velocity tail of the VDF.

This work is organised as follows. We begin with a discussion of the DM velocity distributions, contrasting the SHM with the Gaia-informed distribution. We then describe the inelastic DM scattering framework and the analysis of the LZ230616 event. We present the resulting constraints for the endothermic scattering, and examine the impact of the high-velocity tail, including the possible modification induced by the LMC. We finally discuss broader implications of astrophysical uncertainties for the interpretation of high-energy nuclear recoils in direct-detection experiments.

\noindent \textbf{\emph{Dark matter velocity distribution}--} The event rate in a direct detection experiment depends on the local DM
density $\rho_\chi$, which multiplies the rate, and the velocity distribution of the DM particles that enters through the mean inverse
speed,
\begin{equation}
  \eta(\vmin) = \int_{|\vect{v}|>\vmin} \frac{f_{\rm lab}(\vect{v})}{v}\,{\rm d}^3v,
  \label{eq:eta}
\end{equation}
where $f_{\rm lab}$ is the DM velocity distribution function (VDF) in the frame
of the detector, normalised to unity, and $\vmin$ is the smallest speed that
can produce a recoil of the observed energy. The kinematics that fix $\vmin$,
and the rate itself, are given below. Two features of
Eq.~\eqref{eq:eta} matter here. First, $\eta$ falls as $\vmin$ grows and
vanishes once $\vmin$ exceeds the largest speed available in the laboratory,
$v_{\rm cut}=\vesc+v_{\rm lab}$, where $\vesc$ is the escape speed from the
galaxy at the position of the Sun and $v_{\rm lab}$ is the speed of the
detector through the halo. Second, endothermic scattering needs a large
$\vmin$, so it is fed only by the fastest particles in the halo, which are also
the least well measured. We neglect the annual modulation of $f_{\rm lab}$ due
to the Earth's orbit, a few per cent effect, and use a time-averaged
$v_{\rm lab}$.

\medskip 
\noindent \emph{The standard halo model and its assumptions -- } Direct-detection results are almost always quoted for the Standard Halo Model
(SHM): an isotropic Maxwell-Boltzmann distribution of speeds with most
probable speed $v_0$, truncated at $\vesc$, with a fixed local density
\cite{Drukier1986,LewinSmith1996}. The community convention is
$v_0=238\kms$, $\vesc=544\kms$ and
$\rho_\chi=0.3\GeV\,{\rm cm^{-3}}$ \cite{Baxter2021}.

These three numbers come from three unrelated measurements, and nothing forces
them to describe the same galaxy. The escape speed in particular is measured
from the high-velocity tail of nearby stars, and published values range from
about $500$ to $600\kms$ depending on the sample and the method
\cite{Smith2007,Piffl2014,Deason2019tail}. The Maxwellian shape is also
an assumption rather than a measurement, since the MW has accreted satellites
whose debris is not phase-mixed, and both simulations and Gaia data
show departures from a Maxwellian, mostly in the tail
\cite{Green2017,Evans2019}.

\medskip 
\noindent \emph{A halo model built from Galactic data -- }
We take a different route. Stars and DM move in the same gravitational
potential, even though their velocity distributions differ. We, therefore, use
stars only to measure the potential of the galaxy, and then ask what velocity
distribution the DM must have to sit in that potential in a steady state. The
escape speed, the local density and the shape of the distribution then all come from a single fit to one dataset. The construction has four steps which are summarised below:

\begin{figure}[t]
  \centering
  \includegraphics[width=\columnwidth]{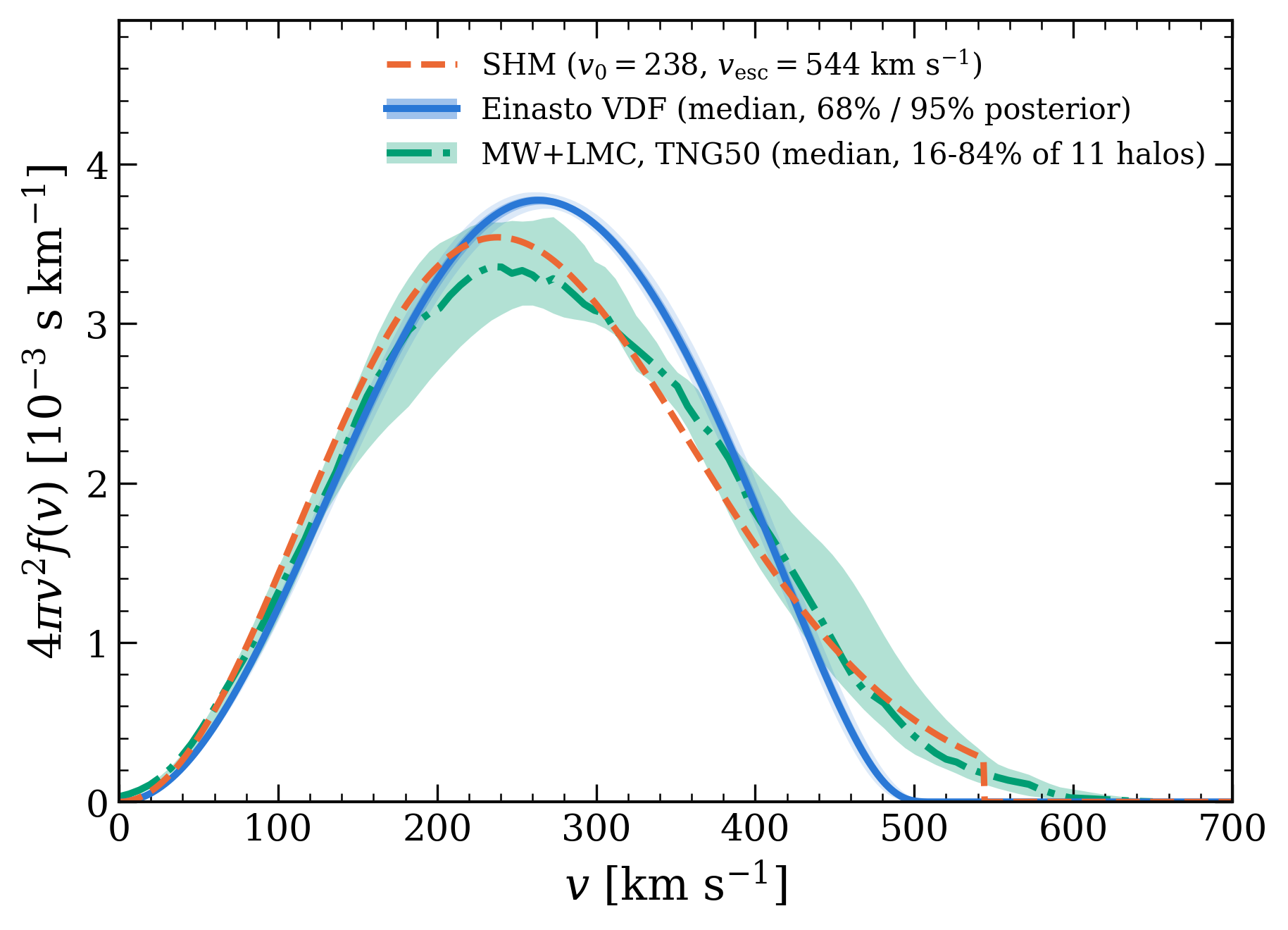}
  \caption{Local DM speed distribution in the Galactic rest frame. Blue: the data-driven distribution obtained by Eddington inversion of the best-fitting Einasto mass model of the Milky Way (median, with 68\% and 95\% bands from 6000 posterior draws). Orange dashed: the Standard Halo Model with $v_0 = 238\kms$ and $\vesc = 544\kms$. Green dash-dotted: the distribution from MW+LMC analogue halos in the TNG50 simulation (median, 16-84\% band of 11 halos).}
  \label{fig:vdf}
\end{figure}

\textit{(i) Stellar catalogue:} We built a catalogue of $3.3\times10^7$ stars
with full six-dimensional phase-space information (position, distance, proper
motion and line-of-sight velocity), by cross-matching Gaia~DR3
astrometry \cite{Gaia2016,GaiaDR3} with 14 spectroscopic and
variable-star surveys, on a common distance and velocity scale
\cite{Mukherjee2026cat}. It covers Galactocentric radii $R$ from the inner
disc to $\sim250$\,kpc and includes $5.2\times10^5$ halo tracers.

\textit{(ii) Rotation curve:} From this catalogue we measured the circular
speed $V_c(R)$, the speed of a circular orbit at radius $R$, which is set by
the mass enclosed within $R$. We used the Jeans equations, which relate the
mean motions and the velocity dispersions of a tracer population to the
underlying potential, between 5 and 250\,kpc, and gas kinematics from the
literature inside 5\,kpc \cite{Bhattacharjee2014,Sofue2020,Mukherjee2026rc}.

\textit{(iii) Mass model:} We fitted the resulting curve with a grid of
36 halo-plus-baryon models (four halo profiles, nine descriptions of the discs
and bulge), at four choices of the Sun's Galactocentric distance $R_0$ and
local circular speed $V_0$, using Markov-chain Monte Carlo sampling
\cite{ForemanMackey2013,Srivastav2026mass}. The data prefer an Einasto
halo \cite{Einasto1965} with a thin and a thick stellar disc, a gas disc and a
bulge. The Einasto profile has a logarithmic density slope
${\rm d}\ln\rho_\chi/{\rm d}\ln r = -2(r/r_s)^{\alpha}$, with a scale radius
$r_s$ and a shape index $\alpha$ that the fit is free to choose. The
preferred fit has $\alpha\simeq0.49$ and a virial mass
$M_{200}\simeq4\times10^{11}M_\odot$, the mass inside the radius where the mean
enclosed density is 200 times the critical density of the Universe.

\textit{(iv) Eddington inversion:} For a spherical halo with an isotropic and
time-independent velocity distribution, the phase-space distribution
$f(\mathcal{E})$ is fixed uniquely by the DM density and the total potential
through Eddington's formula \cite{Eddington1916,BinneyTremaine2008},
\begin{equation}
  f(\mathcal{E}) = \frac{1}{\sqrt{8}\,\pi^2}
  \int_0^{\mathcal{E}} \frac{{\rm d}^2\rho_\chi}{{\rm d}\Psi^2}
  \frac{{\rm d}\Psi}{\sqrt{\mathcal{E}-\Psi}}.
  \label{eq:eddington}
\end{equation}
Here $\Psi(r)=\int_r^\infty V_c^2(r')\,{\rm d}r'/r'$ is the potential of the
halo \emph{and} the baryons, defined to vanish at infinity, and
$\mathcal{E}=\Psi-v^2/2$ is the corresponding energy per unit mass, so that
particles with $\mathcal{E}>0$ are bound. The DM density enters as the tracer,
the potential includes every fitted mass component, and the speed distribution
at the solar radius follows,
\begin{equation}
  g(v) \propto 4\pi v^2 f\!\left(\Psi(R_0)-v^2/2\right),
  \qquad 0\le v\le \vesc,
  \label{eq:gv}
\end{equation}
normalised to unity, with $\vesc=\sqrt{2\Psi(R_0)}$. The escape speed is
therefore arises from the same potential that determines the distribution's shape and local density.

\noindent \emph{Results} -- Fig.\,\ref{fig:vdf} shows $4\pi v^2f(v)$, and Table~\ref{tab:vdf-inputs} lists the halo inputs used in the rest of this paper. We propagated the uncertainty of the mass model by repeating the inversion for 6000 draws from its posterior, and the bands in Fig.~\ref{fig:vdf} are the 68\% and 95\% ranges. The fit gives
\begin{align}
  \vesc &= 521.5 \pm 2.9\kms, \nonumber\\
  \rho_\chi(R_0) &= 0.56^{+0.02}_{-0.03}\GeV\,{\rm cm^{-3}},
\end{align}
with a local circular speed $V_c(R_0)=234.7^{+1.6}_{-1.7}\kms$.

Our distribution and the SHM have the same mean speed, $\simeq262\kms$, but our peak is higher ($264$ against $238\kms$) and is narrower, so it holds far fewer fast
particles ($1.2\%$ of the DM moves faster than $450\kms$, against $5.3\%$ in
the SHM). It also reaches zero smoothly at an escape speed $22\kms$ lower,
whereas the SHM is cut off abruptly while still at $8\%$ of its peak height. 

To evaluate Eq.~\eqref{eq:eta} we boost this distribution into the frame of the
detector. The Earth moves through the halo at
$\vect{v}_{\rm lab}=(0,V_c(R_0),0)+\vect{v}_{\odot,\rm pec}$, with the Sun's
peculiar motion relative to a circular orbit
$\vect{v}_{\odot,\rm pec}=(9,12,7)\kms$. This gives $v_{\rm lab}=247\kms$ for
our halo and $250\kms$ for the SHM, and hence
$v_{\rm cut}=768.5$ and $794.3\kms$ respectively. A smaller $v_{\rm cut}$
directly limits how large an endothermic mass splitting the LZ event can tolerate.

\begin{table}[t]
  \caption{Local halo quantities for the two models used in this work. Speeds
  are in $\kms$ and densities in $\GeV\,{\rm cm^{-3}}$. The data-driven values
  are medians with 68\% ranges from the mass-model posterior.}
  \label{tab:vdf-inputs}
  \centering
  \small
  \begin{ruledtabular}
  \begin{tabular}{lcc}
    Quantity & Einasto VDF & SHM \\
    \hline
    $v_0$ or $V_c(R_0)\,[\kms]$        & $234.7^{+1.6}_{-1.7}$   & $238$ \\
    $\vesc\,[\kms]$                    & $521.5\pm2.9$           & 544 \\
    $v_{\rm lab}\,[\kms]$              & 247                     & 250 \\
    Mean speed\,$[\kms]$                 & 262                     & 264 \\
    $\rho_\chi(R_0)\,{\rm [GeV\, cm^{-3}]}$            & $0.56^{+0.02}_{-0.03}$  & 0.3 \\
    Most probable speed\,$[\kms]$             & 264                     & 238 \\
    Fraction with $v>450$      & 1.2\%                   & 5.3\% \\
    Edge at $\vesc$            & smooth                  & sharp \\
  \end{tabular}
  \end{ruledtabular}
\end{table}
 \noindent\textit{Effect of the Large Magellanic Cloud--} In recent years, the gravitational effect of the Large Magellanic Cloud (LMC) on the MW DM halo has been studied extensively using simulations\,\cite{Besla:2019xbx,Garavito-Camargo:2019kxw,2020MNRAS.494L..11P,2021Natur.592..534C,2021ApJ...919..109G,2022MNRAS.513L..46D,Smith-Orlik:2023kyl}. It has been shown that the high speed tail of the VDF of DM in the solar neighborhood may be affected due to the proximity of the LMC. It is thought to have experienced its first closest approach to the MW only $\sim 50\,$Million years ago at a distance of $\sim48\,\mathrm{kpc}$\,\cite{Besla:2019xbx}. Such encounter may have accelerated the DM particles in the MW halo as well as mixed particles from both halos\,\cite{Smith-Orlik:2023kyl}. Therefore, it is important to consider VDFs that incorporates extended high velocity end due to such events. The effect of the LMC encounter on the DM distribution in the solar neighborhood was calculated recently in Ref.\,\cite{Folsom:2025lly}. It shows a scatter in the cutoff velocity that may reach $\sim640\kms$, which is markedly higher than the SHM. In Fig.\,\ref{fig:vdf}, we show the VDF inferred from the TNG simulation following Ref.\,\cite{Folsom:2025lly}.

\noindent \textbf{\emph{Sensitivity to the tail of the VDF}--} To demonstrate the effect of the precise shape of the tail of the VDF, we consider a spin-independent DM-nuclear interaction. The nuclear scattering rate spectrum can be written as
\begin{align}
  \frac{{\rm d}R}{{\rm d}E_R}
  = \frac{\rho_\chi\,\sigma_p}{2\,\mchi\,\mu_{\chi p}^2\,\langle m_N\rangle}
    \sum_i x_i\, m_i\, &A_i^2 F_i^2(E_R) \nonumber\\
    &\times \eta\big(v_{{\rm min},i}(E_R)\big)\,.
  \label{eq:rate}
\end{align}
Here, $dR/dE_R$ denotes events/(kg $\cdot$ day $\cdot$ keV), $\sigma_p$ is the DM-proton cross-section, $\mu_{\chi p}$ the dark-matter-proton reduced mass, $x_i, m_i, A_i$ are the number fraction, mass, and mass number of various Xe isotopes present in the detector, $F_i(E_R)$ are the corresponding nuclear form factors, and $\langle m_N\rangle=122\GeV$ is the average nucleus mass. Finally, $\eta$ is the mean inverse speed defined in Eq.\eqref{eq:eta}.

We use the publicly available code \texttt{WimPyDD} to calculate the event rate in Xe\,\cite{Jeong:2021bpl}. Instead of the usual Helm form factor, we use the isoscalar nuclear response $W_M^{00}(q)$, calculated using the shell model and as implemented in {\tt WimPyDD}. It is normalised so that $A_i^2F_i^2 \equiv 16\pi\,W_{M,i}^{00}(q)/(2j_i+1)$, where $j_i$ is the nuclear spin. The choice matters as xenon nucleus has a second diffraction minimum is at $267\keV$, inside the $1\sigma$ range of the event.

A recoil of energy $E_R$ on an isotope $i$ requires a minimum
DM speed
\begin{equation}
  v_{{\rm min},i}(E_R) = \frac{1}{\sqrt{2 m_i E_R}}
  \left| \frac{m_i E_R}{\mu_{\chi i}} + \delta \right|,
  \label{eq:vmin}
\end{equation}
\begin{figure*}[ht]
  \centering
    \includegraphics[width=0.95\columnwidth]{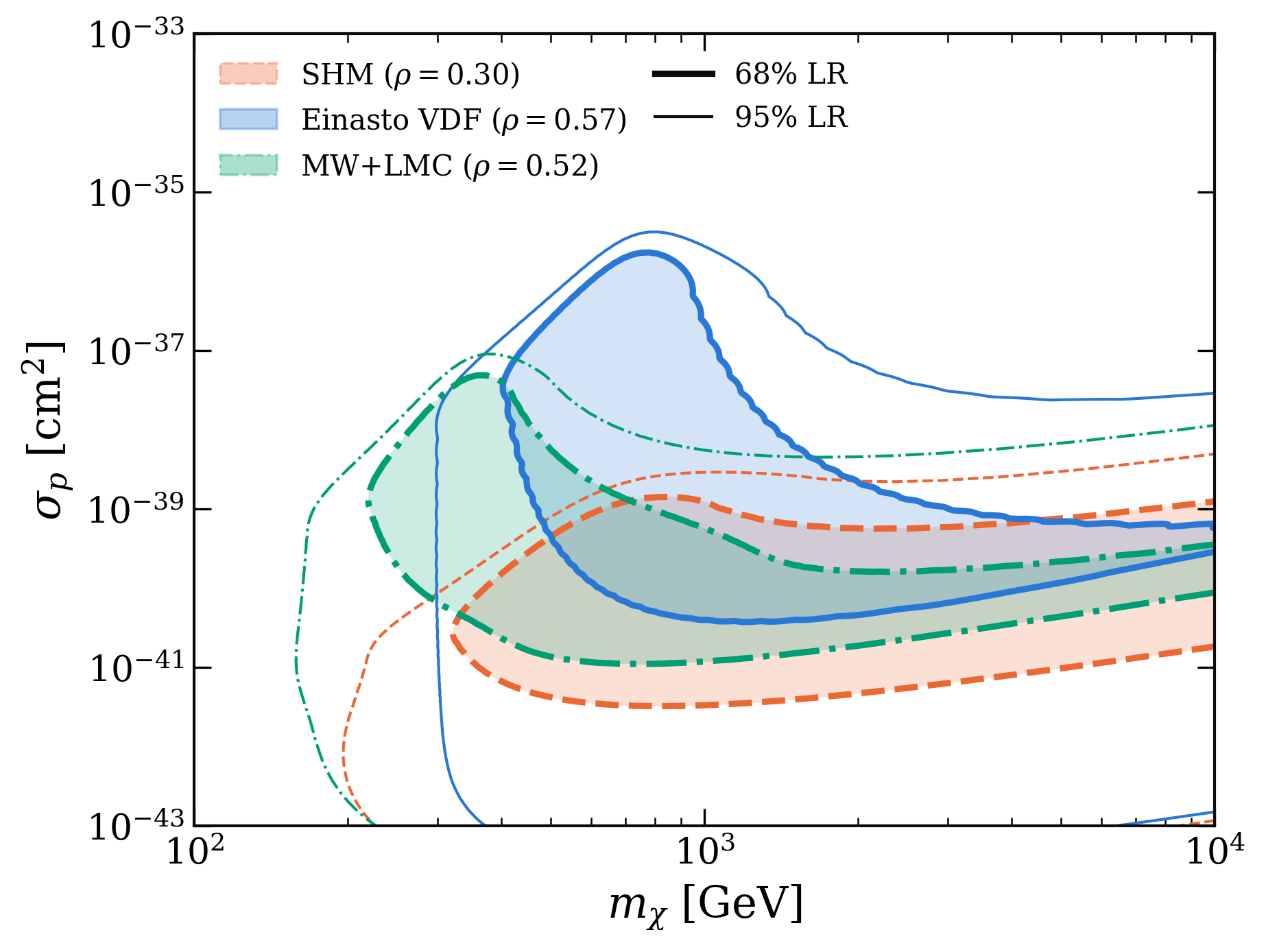}~\includegraphics[width=0.92\columnwidth]{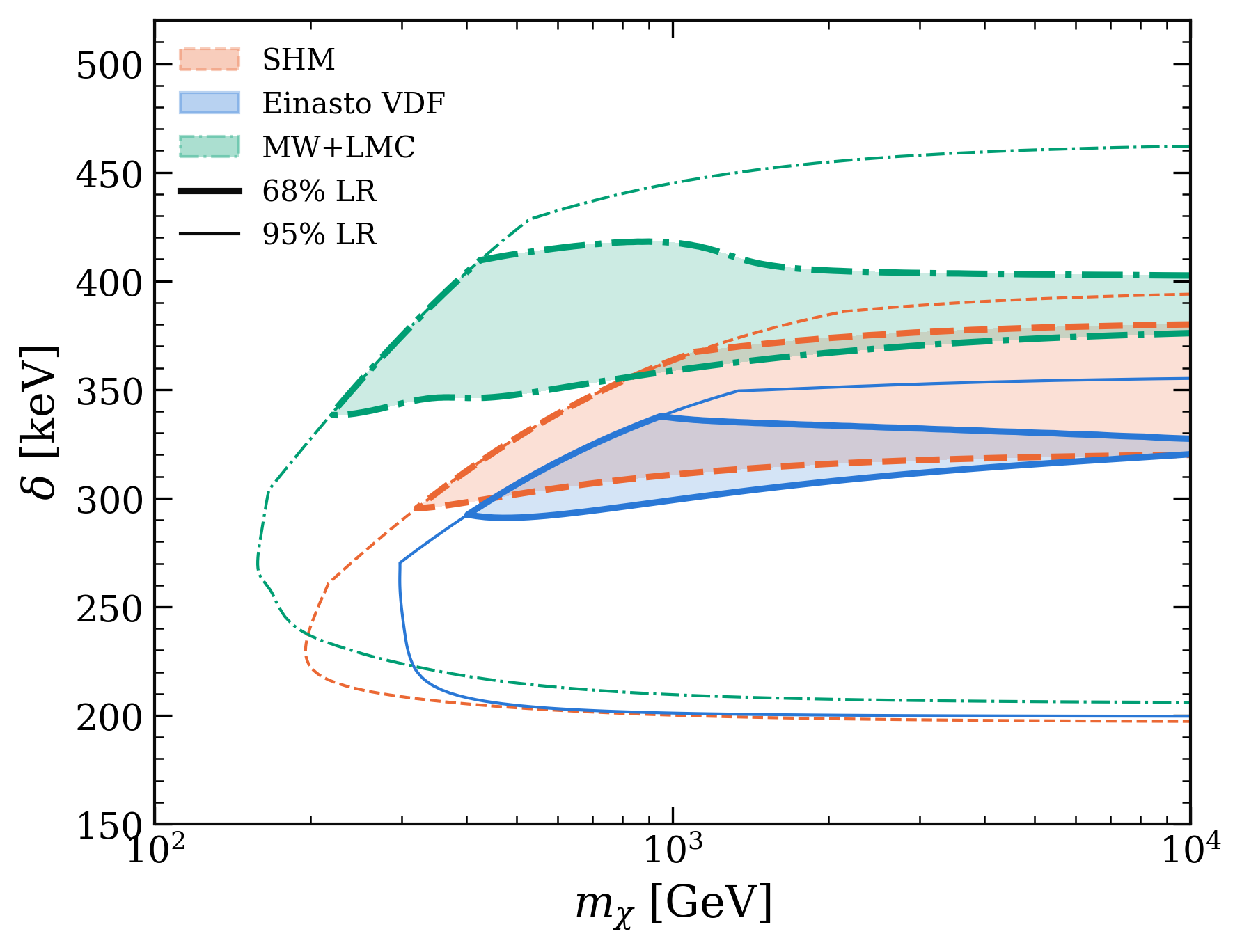}
  \caption{The best-fit regions- 68\% and 95\% likelihood ratio (LR) in $m_\chi-\sigma_p$ (\emph{Left}) and $\mchi-\delta$ (\emph{Right}) for endothermic scattering. Three colors correspond to the SHM (\emph{orange}), the Einasto VDF (\emph{blue}), and SHM+LMC (\emph{green}). The solar neighborhood DM density is shown in the legend.}
  \label{fig:regions_endo}
\end{figure*}
with $\mu_{\chi i}$ the DM-nucleus reduced mass. This minimum speed needs to be smaller than $v_{\rm cut} \equiv \vesc + v_{\rm lab}$ in the lab, where $\vesc$ is the Galactic escape speed at the Sun and $v_{\rm lab}$ the speed of the laboratory relative to the galactic rest frame. Two possibilities emerge depending on the nature of the reaction (below, $m_N$ and $\mu_{\chi N}$ refer to any one of the xenon isotopes):
\begin{itemize}
\item \textit{Endothermic edge:} For $\delta>0$, the lighter DM state needs to have a minimum velocity to upscatter to the heavier state. The smallest possible $v_{\rm min}$ in this case is $v_* = \sqrt{2\delta/\mu_{\chi N}}$, reached at  $E_* = \mu_{\chi N}\delta/m_N$. An event is possible only if $v_* < v_{\rm cut}$, i.e.,
  \begin{equation}
    \delta < \delta_{\rm edge} \equiv \frac{1}{2}\mu_{\chi N} v_{\rm cut}^2\,.
    \label{eq:dedge}
  \end{equation}
  As $\delta\to\delta_{\rm edge}$, the spectrum shrinks to a narrow sliver at $E_*$ and the rate approaches zero.

  \item \textit{Exothermic peak:} For $\delta<0$, there is no minimum velocity required. The recoil energy $E_0$ for which $v_{\rm min}=0$ is given by
  \begin{equation}
    E_0 = \frac{m_\chi |\delta|}{m_\chi+m_N},
    \label{eq:E0}
  \end{equation}

\end{itemize}
In the endothermic case, Eq.\eqref{eq:dedge} shows the tight relation between $\delta$ and $v_\mathrm{cut}$. This is absent in the exothermic scenario.

\noindent \textbf{\emph{Effects on the dark matter parameters}--} We fit the \lz{} event reported in Ref.\,\cite{LZ:2026axp} using spin-independent isoscalar interaction cross section between DM and nucleus. This choice is the most simple one as our goal is to show the changes due to the Gaia-inferred velocity distribution. We divide the LZ search region in recoil energy $E_R$ into three parts based on the energy uncertainty of \lz{} and other assumptions which we elaborate below. LZ collaboration reported the reconstructed nuclear recoil energy of the event as $E_R=248 \pm 23\, ({\rm stat}) \pm 23\, ({\rm sys})\keV$. Therefore, we define the range $248 \pm \sqrt{23^2+23^2}\keV$, i.e., $215.5<E_R<280.5\keV$ as W. In Ref.\,\cite{LZ:2026axp}, the search region extends from $E_R\simeq 5\keV$ to $250\keV$. However, they did not report any candidate event in the energy band below $E_R=215.5\keV$. Hence, we define a lower energy sideband S as $5.4<E_R<215.5\keV$.  Finally, we include a higher energy sideband H: $350<E_R<650\keV$. The LZ collaboration did not publish the efficiency in this energy range and use it only for background study. In this work, we extrapolate their efficiency $\epsilon=0.96$ from lower energy range. These energy ranges with the observed number of events are shown below.
\begin{align*}
  {\rm S} &= [5.4,\ 215.5]\keV & &\text{0 events,}\\
  {\rm W} &= [215.5,\ 280.5]\keV & &\text{1 event,}\\
  {\rm H} &= [350,\ 650]\keV & &\text{0 events,}
\end{align*}
We use the following analytic form of the efficiency function in the whole range.

The expected number of events in a region $X$ from DM scattering in time $T$ and detector mass $M$ is
\begin{equation}
  N_X = MT\int_X {\rm d}E_R\ \epsilon_X(E_R)\,\frac{{\rm d}R}{{\rm d}E_R}.
\end{equation}
The probability of observing $n$ events when $N$ are expected is given by the Poisson distribution ${\rm Pois}(n|N_X) = N_X^n e^{-N_X}/n!$. We define the total likelihood as a product over the three regions,
\begin{align}
  \mathcal{L} &= {\rm Pois}(0|N_{\rm S})\ {\rm Pois}(1|N_{\rm W})\
                  {\rm Pois}(0|N_{\rm H}) \nonumber\\
              &= N_{\rm W}\, e^{-(N_{\rm S}+N_{\rm W}+N_{\rm H})}.
  \label{eq:likelihood}
\end{align}
We vary the DM parameters $\mchi, \delta, \sigma_p$ in order to minimise the negative log-likelihood $-2\ln\mathcal{L}$ and find the best-fit point, and use $(2\ln\mathcal{L}_0 - 2\ln\mathcal{L})=2.30$ and 5.99 to draw the 68\% and 95\% likelihood ratio (LR) regions, respectively. We neglect the background in this analysis.

In Fig.\,\ref{fig:regions_endo}, we show these regions in the $\mchi-\sigma_p$ (left panel) and $\mchi-\delta$ (right panel) planes for endothermic DM with SHM and Einasto VDF. In both parameter planes, there is a significant change in the 68\% LR regions between the two VDF choices. The Einasto model prefers larger ($\sim100$ times) $\sigma_p$ as its high-velocity tail is relatively suppressed. This effect is more prominent for $\mchi\lesssim 1\TeV$ as these masses are close to threshold for the detected event and only the DM particles near the tail of the distribution are kinematically allowed to deposit such large energy. For large $\mchi$, the regions depict a $\sigma_p\propto\mchi$ behavior due the flux of the incoming DM particles. In the $\mchi-\delta$ plane, the best-fit regions have a ceiling as predicted by kinematic edge from Eq.\eqref{eq:dedge}. Since the Einasto VDF has lower $v_\mathrm{cut}$, it prefers smaller mass-split $\delta$ compared to the SHM case. This is evident from the right panel of Fig.\,\ref{fig:regions_endo}. However, both models have similar lower limit on $\delta$. For $\mchi\gtrsim 1\TeV$, the reduced mass $\mu_{\chi N}\to m_N$. Hence, the best-fit regions do not have much dependence on $\mchi$. Furthermore, we find that the exothermic scenario does not change appreciably with change in the VDF (see the Appendix for the corresponding results).

We repeat our analysis using the upper edge, i.e., the largest possible $v_{\rm cut}$, of the scatter in the VDF due to the interaction with a MW+LMC model taken from Ref.\,\cite{Folsom:2025lly}\footnote{Note that the actual MW+LMC distribution might be anisotropic. However, we make the simplistic assumption of taking $v_{\rm cut}$ to be similar to that of an isotropic distribution.}. We show the 68\% and 95\% likelihood ratio best-fit results as the green-shaded regions in Fig.\,\ref{fig:regions_endo}. Since the MW+LMC velocity distribution has more particles with higher speeds, even lighter DM particles can give rise to the \lz{} event. This is evident from Fig.\,\ref{fig:regions_endo}. The green contours extend to lower $\mchi$ compared to the other two VDF models. The MW+LMC model also prefers higher $\delta$ as can be seen from the right panel of Fig.\,\ref{fig:regions_endo}. This directly follows from Eq.\eqref{eq:dedge} as the $v_{\rm cut}$ is higher in this case and explains the upward shift of the green contours in $\delta$. However, larger $\delta$ pushes the spectrum toward the high energy sideband H and decreases the overall scattering rate. To compensate for that, the bestfit $\sigma_p$ is slightly greater in this case relative to SHM. This is seen in the left panel of Fig.\,\ref{fig:regions_endo}.

\noindent \textbf{\emph{Summary and Discussions}--} We have investigated the interpretation of the high energy nuclear recoil event, LZ230616, in the framework of inelastic DM, replacing the SHM with a velocity distribution obtained from a Gaia informed model of the MW. The latter is constructed from the observed galactic mass distribution and Eddington inversion,
obtaining a lower escape speed, $\vesc = 521.5 \pm 2.9\kms,$ compared with $\vesc = 544\kms$ in the SHM. This leads to a substantially reduced population of DM in the high velocity tail. For example, we find that Gaia data allows for only $1.2\%$ DM particles with speed above $450\kms$, compared with $5.3\%$ for a SHM (see Table\, \ref{tab:vdf-inputs} for further details).

This leads to a substantial impact on the endothermic DM interpretation of the LZ event, where the recoil energy is intrinsically tied to the maximum available DM speed. For the Gaia informed distribution, the lower laboratory frame cut-off, $v_{cut}= 768.5\kms$ restricts the kinematically accessible mass splitting, shifting it to smaller values as compared with the SHM. Correspondingly, the inferred cross-section is larger by one or two orders of magnitude, especially for masses of DM less than 1 TeV.
Additionally, we have also included studies of the gravitational effect of the LMC on the MW DM halo. Such interactions can accelerate the local DM particles and extend the high velocity tail, substantially above the SHM value. We found that including the LMC with MW extends the parameter space to higher mass-splittings and lower DM masses, than the Gaia only results. In contrast, the difference in velocity distribution has little impact on the exothermic interpretation, where the released energy allows for the observed recoil without requiring high velocity DM particles.

These results demonstrate that the interpretation of high energy recoil events in direct detection experiments can depend sensitively on the underlying astrophysical inputs, particularly, when the effects probe the extreme tail of the DM distribution. In particular, an endothermic interpretation cannot be \emph{unentangled} from a galactic dynamical model used to determine the available DM high velocity phase space distribution. The Gaia informed reconstruction of the phase space distribution provides exactly this self-consistent connection.

There are, of course, a few caveats to be careful about. Our Gaia informed VDF is subject to uncertainties in the assumptions underlying the Eddington inversion process - such as spherical symmetry, velocity isotropy and dynamical equilibrium. While comparisons with hydrodynamic simulations reproduce bulk of the local speed distributions, the detailed shape of the high velocity tail remains model dependent. Furthermore, note  that in the past history of the MW halo, in particular, mergers of smaller DM halos will create localised non-virialised structures\,\cite{Folsom:2025lly}. These may have significantly different velocity distribution than SHM which assumes perfect virialised system. Because the MW halo has gone through such events in the past, the non-virialised VDF components may affect our results.

Overall, this work demonstrates that astrophysical uncertainties are not merely a normalisation effect - they can qualitatively reshape the DM parameter space, especially when the signal relies on the high velocity tail. A careful modeling of the galactic dynamics and the gravitational effects of nearby objects are, therefore, crucial for the DM interpretation of the high energy recoil event at LZ, and in general, any future direct detection experiment. This is true for inelastic scattering of DM as well as for determining the lower mass threshold of the sensitivity of any ongoing and future experiments. This is particularly important for the next-generation experimental proposals that are trying to lower the threshold for DM mass that can be probed\,\cite{EDELWEISS:2020fxc,TESSERACT:2025tfw,Gao:2024irf,Baudis:2025zyn,Schwemmbauer:2025evp,SPICE:2023tru,Maris:2017xvi,Das:2022srn,Das:2024jdz,Das:2023cbv,Sherpa:2026tgy,Arias:2026vog,Delicato:2023wrg,Abbamonte:2025guf}. As these technologies develop, measurement of the high velocity tail of the DM velocity distribution in the solar neighborhood will become important. Future large nuclear recoil energy measurements will also probe this poorly constrained high velocity tail of the DM distribution. Incorporating increasingly precise stellar-kinematic data and realistic models of galactic substructure will therefore be important for connecting future direct detection observations to the underlying DM particle properties.

\bigskip 

\noindent\textit{Acknowledgment--} AD acknowledges ANRF for the financial support through PMECRG (Grant no. ANRF/ECRG/2025/001012/PMS).  MS acknowledges support from the Early Career Research Grant by Anusandhan National Research Foundation (ANRF/ECRG/2024/000522/PMS). We acknowledge the use of Claude for writing efficient codes for this work.

\bibliography{references} 

@ARTICLE{Bhattacharjee2014,
  author = {{Bhattacharjee}, P. and {Chaudhury}, S. and {Kundu}, S.},
  title = {{Rotation Curve of the Milky Way Out to ~200 kpc}},
  journal = {ApJ},
  year = 2014,
  volume = {785},
  pages = {63},
  doi = {10.1088/0004-637X/785/1/63}
}

@ARTICLE{Sofue2020,
  author = {{Sofue}, Y.},
  title = {{Rotation Curve of the Milky Way and the Dark Matter Density}},
  journal = {Galaxies},
  year = 2020,
  volume = {8},
  number = {2},
  pages = {37},
  doi = {10.3390/galaxies8020037}
}

@ARTICLE{Dutton2014,
  author = {{Dutton}, A.~A. and {Macci{\`o}}, A.~V.},
  title = {{Cold dark matter haloes in the Planck era: evolution of structural parameters for Einasto and NFW profiles}},
  journal = {MNRAS},
  year = 2014,
  volume = {441},
  pages = {3359},
  doi = {10.1093/mnras/stu742}
}

@ARTICLE{Eilers2019,
  author = {{Eilers}, A.-C. and {Hogg}, D.~W. and {Rix}, H.-W. and {Ness}, M.~K.},
  title = {{The Circular Velocity Curve of the Milky Way from 5 to 25 kpc}},
  journal = {ApJ},
  year = 2019,
  volume = {871},
  pages = {120},
  doi = {10.3847/1538-4357/aaf648}
}

@ARTICLE{GRAVITY2018,
  author = {{GRAVITY Collaboration} and {Abuter}, R. and {Amorim}, A. and {Anugu}, N. and et al.},
  title = {{Detection of the gravitational redshift in the orbit of the star S2 near the Galactic centre massive black hole}},
  journal = {A\&A},
  year = 2018,
  volume = {615},
  pages = {L15},
  doi = {10.1051/0004-6361/201833718}
}

@ARTICLE{KerrLyndenBell1986,
  author = {{Kerr}, F.~J. and {Lynden-Bell}, D.},
  title = {{Review of galactic constants}},
  journal = {MNRAS},
  year = 1986,
  volume = {221},
  pages = {1023},
  doi = {10.1093/mnras/221.4.1023}
}

@ARTICLE{Navarro1997,
  author = {{Navarro}, J.~F. and {Frenk}, C.~S. and {White}, S.~D.~M.},
  title = {{A Universal Density Profile from Hierarchical Clustering}},
  journal = {ApJ},
  year = 1997,
  volume = {490},
  pages = {493},
  doi = {10.1086/304888}
}

@ARTICLE{Burkert1995,
  author = {{Burkert}, A.},
  title = {{The Structure of Dark Matter Halos in Dwarf Galaxies}},
  journal = {ApJ},
  year = 1995,
  volume = {447},
  pages = {L25},
  doi = {10.1086/309560}
}

@ARTICLE{Einasto1965,
  author = {{Einasto}, J.},
  title = {{Kinematics and dynamics of stellar systems}},
  journal = {Trudy Inst. Astrofiz. Alma-Ata},
  year = 1965,
  volume = {5},
  pages = {87}
}

@ARTICLE{Hernquist1990,
  author = {{Hernquist}, L.},
  title = {{An Analytical Model for Spherical Galaxies and Bulges}},
  journal = {ApJ},
  year = 1990,
  volume = {356},
  pages = {359},
  doi = {10.1086/168845}
}

@ARTICLE{Plummer1911,
  author = {{Plummer}, H.~C.},
  title = {{On the problem of distribution in globular star clusters}},
  journal = {MNRAS},
  year = 1911,
  volume = {71},
  pages = {460},
  doi = {10.1093/mnras/71.5.460}
}

@ARTICLE{Freeman1970,
  author = {{Freeman}, K.~C.},
  title = {{On the Disks of Spiral and S0 Galaxies}},
  journal = {ApJ},
  year = 1970,
  volume = {160},
  pages = {811},
  doi = {10.1086/150474}
}

@ARTICLE{ForemanMackey2013,
  author = {{Foreman-Mackey}, D. and {Hogg}, D.~W. and {Lang}, D. and {Goodman}, J.},
  title = {{emcee: The MCMC Hammer}},
  journal = {PASP},
  year = 2013,
  volume = {125},
  pages = {306},
  doi = {10.1086/670067}
}

@ARTICLE{GelmanRubin1992,
  author = {{Gelman}, A. and {Rubin}, D.~B.},
  title = {{Inference from Iterative Simulation Using Multiple Sequences}},
  journal = {Statistical Science},
  year = 1992,
  volume = {7},
  pages = {457},
  doi = {10.1214/ss/1177011136}
}

@ARTICLE{Watanabe2010,
  author = {{Watanabe}, S.},
  title = {{Asymptotic Equivalence of Bayes Cross Validation and Widely Applicable Information Criterion in Singular Learning Theory}},
  journal = {Journal of Machine Learning Research},
  year = 2010,
  volume = {11},
  pages = {3571}
}

@ARTICLE{Astropy2013,
  author = {{Astropy Collaboration} and {Robitaille}, T.~P. and {Tollerud}, E.~J. and et al.},
  title = {{Astropy: A community Python package for astronomy}},
  journal = {A\&A},
  year = 2013,
  volume = {558},
  pages = {A33},
  doi = {10.1051/0004-6361/201322068}
}

@ARTICLE{Smith2007,
  author = {{Smith}, M.~C. and {Ruchti}, G.~R. and {Helmi}, A. and {Wyse}, R.~F.~G. and et al.},
  title = {{The RAVE survey: constraining the local Galactic escape speed}},
  journal = {MNRAS},
  year = 2007,
  volume = {379},
  pages = {755},
  doi = {10.1111/j.1365-2966.2007.11964.x}
}

@ARTICLE{Xue2008,
  author = {{Xue}, X.~X. and {Rix}, H.~W. and {Zhao}, G. and {Re Fiorentin}, P. and {Naab}, T. and {Steinmetz}, M. and et al.},
  title = {{The Milky Way's Circular Velocity Curve to 60 kpc and an Estimate of the Dark Matter Halo Mass from Kinematics of $\sim$2400 SDSS Blue Horizontal-Branch Stars}},
  journal = {ApJ},
  year = 2008,
  volume = {684},
  pages = {1143},
  doi = {10.1086/589500}
}

@ARTICLE{Piffl2014,
  author = {{Piffl}, T. and {Scannapieco}, C. and {Binney}, J. and {Steinmetz}, M. and et al.},
  title = {{The RAVE survey: the Galactic escape speed and the mass of the Milky Way}},
  journal = {A\&A},
  year = 2014,
  volume = {562},
  pages = {A91},
  doi = {10.1051/0004-6361/201322531}
}

@article{LZ:2026axp,
    author = "Akerib, D. S. and others",
    collaboration = "LZ",
    title = "{Search for dark matter particle interactions in an extended nuclear recoil energy window with the LUX-ZEPLIN (LZ) experiment}",
    eprint = "2609.02823",
    archivePrefix = "arXiv",
    primaryClass = "hep-ex",
    doi = "10.17182/hepdata.182472.v1",
    year = "2026",
    journal = "{arXiv e-prints}"
}

@article{Jeong:2021bpl,
    author = "Jeong, Injun and Kang, Sunghyun and Scopel, Stefano and Tomar, Gaurav",
    title = "{WimPyDD: An object{\textendash}oriented Python code for the calculation of WIMP direct detection signals}",
    eprint = "2106.06207",
    archivePrefix = "arXiv",
    primaryClass = "hep-ph",
    reportNumber = "CQUeST-2021-0663, TUM-HEP 1343/21",
    doi = "10.1016/j.cpc.2022.108342",
    journal = "Comput. Phys. Commun.",
    volume = "276",
    pages = "108342",
    year = "2022"
}

@MISC{Mukherjee2026cat,
  author = {{Mukherjee}, S. and {Srivastav}, A. and {Majumdar}, S.},
  title = {{Mapping the Milky Way in Six Dimensions: A contiguous, homogenised, phase-space catalogue of Gaia DR3 tracers up to 250 kpc}},
  year = 2026,
  eprint = {2609.28652},
  archivePrefix = {arXiv}
}

@UNPUBLISHED{Mukherjee2026rc,
  author = {{Mukherjee}, S. and {Srivastav}, A. and {Majumdar}, S.},
  title = {{The Milky Way rotation curve from 5 to 250 kpc}},
  year = 2026,
  note = {in preparation}
}

@UNPUBLISHED{Srivastav2026mass,
  author = {{Srivastav}, A. and {Mukherjee}, S. and {Majumdar}, S.},
  title = {{Bayesian mass modelling of the Milky Way: smooth halo--disc decomposition and evidence for ring-like substructure in the rotation curve}},
  year = 2026,
  note = {in preparation}
}

@ARTICLE{Gaia2016,
  author = {{Gaia Collaboration} and {Prusti}, T. and et al.},
  title = {{The Gaia mission}},
  journal = {A\&A},
  year = 2016,
  volume = {595},
  pages = {A1},
  doi = {10.1051/0004-6361/201629272}
}

@ARTICLE{GaiaDR3,
  author = {{Gaia Collaboration} and {Vallenari}, A. and et al.},
  title = {{Gaia Data Release 3. Summary of the content and survey properties}},
  journal = {A\&A},
  year = 2023,
  volume = {674},
  pages = {A1},
  doi = {10.1051/0004-6361/202243940}
}

@ARTICLE{Lindegren2021,
  author = {{Lindegren}, L. and et al.},
  title = {{Gaia Early Data Release 3. Parallax bias versus magnitude, colour, and position}},
  journal = {A\&A},
  year = 2021,
  volume = {649},
  pages = {A4},
  doi = {10.1051/0004-6361/202039653}
}

@ARTICLE{DESI_DR1,
  author = {{DESI Collaboration}},
  title = {{Data Release 1 of the Dark Energy Spectroscopic Instrument}},
  journal = {arXiv e-prints},
  year = 2025,
  eprint = {2503.14745},
  archivePrefix = {arXiv}
}

@ARTICLE{SEGUE2009,
  author = {{Yanny}, B. and et al.},
  title = {{SEGUE: A Spectroscopic Survey of 240,000 Stars with g = 14-20}},
  journal = {AJ},
  year = 2009,
  volume = {137},
  pages = {4377},
  doi = {10.1088/0004-6256/137/5/4377}
}

@ARTICLE{APOGEE_DR17,
  author = {{Abdurro'uf} and et al.},
  title = {{The Seventeenth Data Release of the Sloan Digital Sky Surveys: Complete Release of MaNGA, MaStar, and APOGEE-2 Data}},
  journal = {ApJS},
  year = 2022,
  volume = {259},
  pages = {35},
  doi = {10.3847/1538-4365/ac4414}
}

@ARTICLE{LAMOST2012,
  author = {{Cui}, X.-Q. and et al.},
  title = {{The Large Sky Area Multi-Object Fiber Spectroscopic Telescope (LAMOST)}},
  journal = {RAA},
  year = 2012,
  volume = {12},
  pages = {1197},
  doi = {10.1088/1674-4527/12/9/003}
}

@ARTICLE{GALAH_DR4,
  author = {{Buder}, S. and et al.},
  title = {{The GALAH Survey: Data Release 4}},
  journal = {arXiv e-prints},
  year = 2024,
  eprint = {2409.19858},
  archivePrefix = {arXiv}
}

@ARTICLE{GaiaESO2022,
  author = {{Gilmore}, G. and et al.},
  title = {{The Gaia-ESO Public Spectroscopic Survey: Motivation, implementation, GIRAFFE data processing, analysis, and final data products}},
  journal = {A\&A},
  year = 2022,
  volume = {666},
  pages = {A120},
  doi = {10.1051/0004-6361/202243134}
}

@ARTICLE{RAVE_DR6,
  author = {{Steinmetz}, M. and et al.},
  title = {{The Sixth Data Release of the Radial Velocity Experiment (RAVE). I. Survey Description, Spectra, and Radial Velocities}},
  journal = {AJ},
  year = 2020,
  volume = {160},
  pages = {82},
  doi = {10.3847/1538-3881/ab9ab9}
}

@ARTICLE{Hogg2019,
  author = {{Hogg}, D.~W. and {Eilers}, A.-C. and {Rix}, H.-W.},
  title = {{Spectrophotometric Parallaxes with Linear Models: Accurate Distances for Luminous Red-giant Stars}},
  journal = {AJ},
  year = 2019,
  volume = {158},
  pages = {147},
  doi = {10.3847/1538-3881/ab398c}
}

@BOOK{BinneyTremaine2008,
  author = {{Binney}, J. and {Tremaine}, S.},
  title = {{Galactic Dynamics}},
  edition = {2nd},
  publisher = {Princeton University Press},
  address = {Princeton},
  year = 2008
}

@ARTICLE{Eddington1916,
  author = {{Eddington}, A.~S.},
  title = {{The distribution of stars in globular clusters}},
  journal = {MNRAS},
  year = 1916,
  volume = {76},
  pages = {572},
  doi = {10.1093/mnras/76.7.572}
}

@ARTICLE{Widrow2000,
  author = {{Widrow}, L.~M.},
  title = {{Distribution Functions for Cuspy Dark Matter Density Profiles}},
  journal = {ApJS},
  year = 2000,
  volume = {131},
  pages = {39},
  doi = {10.1086/317367}
}

@ARTICLE{Drukier1986,
  author = {{Drukier}, A.~K. and {Freese}, K. and {Spergel}, D.~N.},
  title = {{Detecting cold dark-matter candidates}},
  journal = {Phys. Rev. D},
  year = 1986,
  volume = {33},
  pages = {3495},
  doi = {10.1103/PhysRevD.33.3495}
}

@ARTICLE{LewinSmith1996,
  author = {{Lewin}, J.~D. and {Smith}, P.~F.},
  title = {{Review of mathematics, numerical factors, and corrections for dark matter experiments based on elastic nuclear recoil}},
  journal = {Astropart. Phys.},
  year = 1996,
  volume = {6},
  pages = {87},
  doi = {10.1016/S0927-6505(96)00047-3}
}

@ARTICLE{Baxter2021,
  author = {{Baxter}, D. and et al.},
  title = {{Recommended conventions for reporting results from direct dark matter searches}},
  journal = {Eur. Phys. J. C},
  year = 2021,
  volume = {81},
  pages = {907},
  eprint = {2105.00599},
  archivePrefix = {arXiv},
  doi = {10.1140/epjc/s10052-021-09655-y}
}

@ARTICLE{Green2017,
  author = {{Green}, A.~M.},
  title = {{Astrophysical uncertainties on the local dark matter distribution and direct detection experiments}},
  journal = {J. Phys. G},
  year = 2017,
  volume = {44},
  pages = {084001},
  eprint = {1703.10102},
  archivePrefix = {arXiv},
  doi = {10.1088/1361-6471/aa7819}
}

@ARTICLE{Evans2019,
  author = {{Evans}, N.~W. and {O'Hare}, C.~A.~J. and {McCabe}, C.},
  title = {{Refinement of the standard halo model for dark matter searches in light of the Gaia Sausage}},
  journal = {Phys. Rev. D},
  year = 2019,
  volume = {99},
  pages = {023012},
  eprint = {1810.11468},
  archivePrefix = {arXiv},
  doi = {10.1103/PhysRevD.99.023012}
}

@ARTICLE{Lacroix2020,
    author = {{Lacroix}, T. and {N{\'u}{\~n}ez-Casti{\~n}eyra}, A. and {Stref}, M. and {Lavalle}, J. and {Nezri}, E.},
    title = {{Predicting the dark matter velocity distribution in galactic structures: tests against hydrodynamic cosmological simulations}},
    journal = {JCAP},
    year = 2020,
    volume = {10},
    number = {10},
    pages = {031},
    eprint = {1805.02403},
    archivePrefix = {arXiv},
    doi = {10.1088/1475-7516/2020/10/031}
}

@ARTICLE{Bhattacharjee2013,
  author = {{Bhattacharjee}, P. and {Chaudhury}, S. and {Kundu}, S. and {Majumdar}, S.},
  title = {{Deriving the velocity distribution of Galactic dark matter particles from the rotation curve data}},
  journal = {Phys. Rev. D},
  year = 2013,
  volume = {87},
  pages = {083525},
  eprint = {1210.2328},
  archivePrefix = {arXiv},
  doi = {10.1103/PhysRevD.87.083525}
}

@ARTICLE{Deason2019tail,
  author = {{Deason}, A.~J. and {Fattahi}, A. and {Belokurov}, V. and {Evans}, N.~W. and {Grand}, R.~J.~J. and {Marinacci}, F. and {Pakmor}, R.},
  title = {{The local high-velocity tail and the Galactic escape speed}},
  journal = {MNRAS},
  year = 2019,
  volume = {485},
  pages = {3514},
  eprint = {1901.02016},
  archivePrefix = {arXiv},
  doi = {10.1093/mnras/stz623}
}

@article{Folsom:2025lly,
    author = "Folsom, Dylan and Blanco, Carlos and Lisanti, Mariangela and Necib, Lina and Vogelsberger, Mark and Hernquist, Lars",
    title = "{Dark Matter Velocity Distributions for Direct Detection: Astrophysical Uncertainties Are Smaller Than They Appear}",
    eprint = "2505.07924",
    archivePrefix = "arXiv",
    primaryClass = "hep-ph",
    doi = "10.1103/wmpq-mw4h",
    journal = "Phys. Rev. Lett.",
    volume = "135",
    number = "21",
    pages = "211004",
    year = "2025"
}

@article{Smith-Orlik:2023kyl,
    author = "Smith-Orlik, Adam and others",
    title = "{The impact of the Large Magellanic Cloud on dark matter direct detection signals}",
    eprint = "2302.04281",
    archivePrefix = "arXiv",
    primaryClass = "astro-ph.GA",
    doi = "10.1088/1475-7516/2023/10/070",
    journal = "JCAP",
    volume = "10",
    number = "10",
    pages = "070",
    year = "2023"
}

@article{Besla:2019xbx,
    author = "Besla, Gurtina and Peter, Annika and Garavito-Camargo, Nicolas",
    title = "{The highest-speed local dark matter particles come from the Large Magellanic Cloud}",
    eprint = "1909.04140",
    archivePrefix = "arXiv",
    primaryClass = "astro-ph.GA",
    doi = "10.1088/1475-7516/2019/11/013",
    journal = "JCAP",
    volume = "11",
    number = "11",
    pages = "013",
    year = "2019"
}

@ARTICLE{2022MNRAS.513L..46D,
       author = {{Donaldson}, Katelin and {Petersen}, Michael S. and {Pe{\~n}arrubia}, Jorge},
        title = "{Effects on the local dark matter distribution due to the large magellanic cloud}",
      journal = {\mnras},
         year = 2022,
        month = jun,
       volume = {513},
       number = {1},
        pages = {46-51},
          doi = {10.1093/mnrasl/slac031},
archivePrefix = {arXiv},
       eprint = {2111.15440},
 primaryClass = {astro-ph.GA},
       adsurl = {https://ui.adsabs.harvard.edu/abs/2022MNRAS.513L..46D}
}

@article{Garavito-Camargo:2019kxw,
    author = "Garavito-Camargo, Nicolas and Besla, Gurtina and Laporte, Chervin F. P. and Johnston, Kathryn V. and G{\'o}mez, Facundo A. and Watkins, Laura L.",
    title = "{Hunting for the Dark Matter Wake Induced by the Large Magellanic Cloud}",
    journal = {ApJ},
    eprint = "1902.05089",
    archivePrefix = "arXiv",
    primaryClass = "astro-ph.GA",
    doi = "10.3847/1538-4357/ab32eb",
    month = "2",
    year = "2019"
}

@ARTICLE{2021ApJ...919..109G,
       author = {{Garavito-Camargo}, Nicol{\'a}s and {Besla}, Gurtina and {Laporte}, Chervin F.~P. and {Price-Whelan}, Adrian M. and {Cunningham}, Emily C. and {Johnston}, Kathryn V. and {Weinberg}, Martin and {G{\'o}mez}, Facundo A.},
        title = "{Quantifying the Impact of the Large Magellanic Cloud on the Structure of the Milky Way's Dark Matter Halo Using Basis Function Expansions}",
      journal = {\apj},
         year = 2021,
        month = oct,
       volume = {919},
       number = {2},
          eid = {109},
        pages = {109},
          doi = {10.3847/1538-4357/ac0b44},
archivePrefix = {arXiv},
       eprint = {2010.00816},
 primaryClass = {astro-ph.GA},
       adsurl = {https://ui.adsabs.harvard.edu/abs/2021ApJ...919..109G}
}

@ARTICLE{2020MNRAS.494L..11P,
       author = {{Petersen}, Michael S. and {Pe{\~n}arrubia}, Jorge},
        title = "{Reflex motion in the Milky Way stellar halo resulting from the Large Magellanic Cloud infall}",
      journal = {\mnras},
         year = 2020,
        month = may,
       volume = {494},
       number = {1},
        pages = {L11-L16},
          doi = {10.1093/mnrasl/slaa029},
archivePrefix = {arXiv},
       eprint = {2001.09142},
 primaryClass = {astro-ph.GA},
       adsurl = {https://ui.adsabs.harvard.edu/abs/2020MNRAS.494L..11P}
}

@ARTICLE{2021Natur.592..534C,
       author = {{Conroy}, Charlie and {Naidu}, Rohan P. and {Garavito-Camargo}, Nicol{\'a}s and {Besla}, Gurtina and {Zaritsky}, Dennis and {Bonaca}, Ana and {Johnson}, Benjamin D.},
        title = "{All-sky dynamical response of the Galactic halo to the Large Magellanic Cloud}",
      journal = {\nat},
         year = 2021,
        month = apr,
       volume = {592},
       number = {7855},
        pages = {534-536},
          doi = {10.1038/s41586-021-03385-7},
archivePrefix = {arXiv},
       eprint = {2104.09515},
 primaryClass = {astro-ph.GA},
       adsurl = {https://ui.adsabs.harvard.edu/abs/2021Natur.592..534C}
}

@article{TESSERACT:2025tfw,
    author = "Bui, T. K. and others",
    collaboration = "TESSERACT",
    title = "{First Limits on Light Dark Matter Interactions in a Low Threshold Two-Channel Athermal Phonon Detector from the TESSERACT Collaboration}",
    eprint = "2503.03683",
    archivePrefix = "arXiv",
    primaryClass = "hep-ex",
    doi = "10.1103/hsrl-crvf",
    journal = "Phys. Rev. Lett.",
    volume = "135",
    number = "16",
    pages = "161002",
    year = "2025"
}

@article{Baudis:2025zyn,
    author = "Baudis, Laura and others",
    title = "{First Sub-MeV Dark Matter Search with the QROCODILE Experiment Using Superconducting Nanowire Single-Photon Detectors}",
    doi = "10.1103/4hb6-f6jl",
    journal = "Phys. Rev. Lett.",
    volume = "135",
    number = "8",
    pages = "081002",
    year = "2025"
}

@article{Das:2023cbv,
    author = "Das, Anirban and Jang, Jiho and Min, Hongki",
    title = "{Sub-MeV dark matter detection with bilayer graphene}",
    eprint = "2312.00866",
    archivePrefix = "arXiv",
    primaryClass = "hep-ph",
    doi = "10.1103/PhysRevD.110.043020",
    journal = "Phys. Rev. D",
    volume = "110",
    number = "4",
    pages = "043020",
    year = "2024"
}

@article{SPICE:2023tru,
    author = "Anthony-Petersen, R. and others",
    collaboration = "SPICE, HeRALD",
    title = "{Demonstration of the HeRALD superfluid helium detector concept}",
    eprint = "2307.11877",
    archivePrefix = "arXiv",
    primaryClass = "physics.ins-det",
    doi = "10.1103/PhysRevD.110.072006",
    journal = "Phys. Rev. D",
    volume = "110",
    number = "7",
    pages = "072006",
    year = "2024"
}

@article{Maris:2017xvi,
    author = "Maris, Humphrey J. and Seidel, George M. and Stein, Derek",
    title = "{Dark Matter Detection Using Helium Evaporation and Field Ionization}",
    eprint = "1706.00117",
    archivePrefix = "arXiv",
    primaryClass = "astro-ph.IM",
    doi = "10.1103/PhysRevLett.119.181303",
    journal = "Phys. Rev. Lett.",
    volume = "119",
    number = "18",
    pages = "181303",
    year = "2017"
}

@article{Das:2022srn,
    author = "Das, Anirban and Kurinsky, Noah and Leane, Rebecca K.",
    title = "{Dark Matter Induced Power in Quantum Devices}",
    eprint = "2210.09313",
    archivePrefix = "arXiv",
    primaryClass = "hep-ph",
    reportNumber = "SLAC-PUB-17691",
    doi = "10.1103/PhysRevLett.132.121801",
    journal = "Phys. Rev. Lett.",
    volume = "132",
    number = "12",
    pages = "121801",
    year = "2024"
}

@article{Das:2024jdz,
    author = "Das, Anirban and Kurinsky, Noah and Leane, Rebecca K.",
    title = "{Transmon Qubit constraints on dark matter-nucleon scattering}",
    eprint = "2405.00112",
    archivePrefix = "arXiv",
    primaryClass = "hep-ph",
    reportNumber = "SLAC-PUB-17769",
    doi = "10.1007/JHEP07(2024)233",
    journal = "JHEP",
    volume = "07",
    number = "07",
    pages = "233",
    year = "2024"
}

@article{Sherpa:2026tgy,
    author = "Sherpa, Rinchen and Sarkar, Anuvab and Maity, Tarak Nath and Dutta, Paramita and Laha, Ranjan and Das, Anirban",
    title = "{Dive deeper with SUBMARINE: SUB-Mev dArk matter diRect detectIon using bilayer grapheNE}",
    eprint = "2604.21969",
    archivePrefix = "arXiv",
    primaryClass = "hep-ph",
    month = "4",
    year = "2026",
    journal = "{arXiv e-prints}"
}

@article{Arias:2026vog,
    author = "Arias, Tom{\'a}s and Bellinvia, Antonino and Cavoto, Gianluca and Esposito, Angelo and Pandolfi, Francesco and Papiri, Guglielmo and Polosa, Antonio D. and Wu, Tyler",
    title = "{Hydrogenated carbon structures as directional sub-GeV dark matter detectors}",
    eprint = "2602.02694",
    archivePrefix = "arXiv",
    primaryClass = "hep-ph",
    doi = "10.1103/ym4j-8zlp",
    journal = "Phys. Rev. D",
    volume = "114",
    number = "1",
    pages = "L011702",
    year = "2026"
}

@article{Abbamonte:2025guf,
    author = "Abbamonte, P. and others",
    title = "{SPLENDOR: a novel detector platform to search for light dark matter with narrow-gap semiconductors}",
    eprint = "2507.17782",
    archivePrefix = "arXiv",
    primaryClass = "physics.ins-det",
    reportNumber = "LA-UR-25-26113",
    month = "7",
    year = "2025",
    journal = "{arXiv e-prints}"
}

@article{Schwemmbauer:2025evp,
    author = "Schwemmbauer, Christina and others",
    title = "{First direct search for light dark matter interactions in a transition-edge sensor}",
    eprint = "2506.18982",
    archivePrefix = "arXiv",
    primaryClass = "physics.ins-det",
    reportNumber = "MIT-CTP/5879, DESY-25-086",
    month = "6",
    year = "2025",
    journal = "{arXiv e-prints}"
}

@article{Gao:2024irf,
    author = "Gao, Jiansong and Hochberg, Yonit and Lehmann, Benjamin V. and Nam, Sae Woo and Szypryt, Paul and Vissers, Michael R. and Xu, Tao",
    title = "{Detecting Light Dark Matter with Kinetic Inductance Detectors}",
    eprint = "2403.19739",
    archivePrefix = "arXiv",
    primaryClass = "hep-ph",
    reportNumber = "MIT-CTP/5654",
    month = "3",
    year = "2024",
    journal = "{arXiv e-prints}"
}

@article{Delicato:2023wrg,
    author = "Delicato, D. and others",
    title = "{Low-energy spectrum of the BULLKID detector array operated on surface}",
    eprint = "2308.14399",
    archivePrefix = "arXiv",
    primaryClass = "hep-ex",
    doi = "10.1140/epjc/s10052-024-12714-9",
    journal = "Eur. Phys. J. C",
    volume = "84",
    number = "4",
    pages = "353",
    year = "2024"
}

@article{EDELWEISS:2020fxc,
    author = "Arnaud, Q. and others",
    collaboration = "EDELWEISS",
    title = "{First germanium-based constraints on sub-MeV Dark Matter with the EDELWEISS experiment}",
    eprint = "2003.01046",
    archivePrefix = "arXiv",
    primaryClass = "astro-ph.GA",
    doi = "10.1103/PhysRevLett.125.141301",
    journal = "Phys. Rev. Lett.",
    volume = "125",
    number = "14",
    pages = "141301",
    year = "2020"
}

@article{Wu:2026nhi,
    author = "Wu, Lei and Zhang, Yang and Zhu, Bin",
    title = "{TeV Higgsino Dark Matter from LZ Nuclear Recoil to Fermi-LAT Gamma Rays}",
    journal = "{arXiv e-prints}",
    eprint = "2609.01590",
    archivePrefix = "arXiv",
    primaryClass = "hep-ph",
    month = "9",
    year = "2026"
}

@article{Freese:2026sga,
    author = "Freese, Katherine and Theodosopoulos, Dionysios P.",
    title = "{Higgsino Dark Matter Interpretation of the LUX-ZEPLIN 248 keV Nuclear-Recoil Event}",
    journal = "{arXiv e-prints}",
    eprint = "2609.01583",
    archivePrefix = "arXiv",
    primaryClass = "hep-ph",
    month = "9",
    year = "2026"
}

@article{Su:2026rwz,
    author = "Su, Liangliang and Yang, Jin Min and Yang, Wen-Na",
    title = "{Inelastic Dark Matter Signature at High Recoil Energy in LUX-ZEPLIN and CRESST}",
    journal = "{arXiv e-prints}",
    eprint = "2609.01475",
    archivePrefix = "arXiv",
    primaryClass = "hep-ph",
    month = "9",
    year = "2026"
}

@article{Yamashita:2026ump,
    author = "Yamashita, Kimiko",
    title = "{Inelastic Dark Photon Dark Matter for the LUX-ZEPLIN High-Recoil Event and the Galactic Halo Gamma-Ray Excess}",
    journal = "{arXiv e-prints}",
    eprint = "2609.02868",
    archivePrefix = "arXiv",
    primaryClass = "hep-ph",
    month = "9",
    year = "2026"
}

@article{Nomura:2026qyq,
    author = "Nomura, Yasunori",
    title = "{Dark Matter as the Z{\_}2 Partner of the Standard Model Higgs Boson}",
    journal = "{arXiv e-prints}",
    eprint = "2609.02505",
    archivePrefix = "arXiv",
    primaryClass = "hep-ph",
    reportNumber = "RIKEN-iTHEMS-Report-26",
    month = "9",
    year = "2026"
}

@article{Visinelli:2026kgt,
    author = "Visinelli, Luca",
    title = "{A Peccei-Quinn Origin for Inelastic Electroweak Dark Matter after LUX-ZEPLIN}",
    journal = "{arXiv e-prints}",
    eprint = "2609.02807",
    archivePrefix = "arXiv",
    primaryClass = "hep-ph",
    month = "9",
    year = "2026"
}

@article{Pospelov:2026ewn,
    author = "Pospelov, Maxim and Ramani, Harikrishnan",
    title = "{Strong Constraints on Higgsino Dark Matter from Solar Capture}",
    journal = "{arXiv e-prints}",
    eprint = "2609.02775",
    archivePrefix = "arXiv",
    primaryClass = "hep-ph",
    month = "9",
    year = "2026"
}

@article{Du:2026guj,
    author = "Du, Xiaokang and Wang, Fei",
    title = "{TeV Higgsino Interpretation of the LZ High-Recoil Event with Intermediate-Scale Electroweak Gauginos}",
    journal = "{arXiv e-prints}",
    eprint = "2609.04163",
    archivePrefix = "arXiv",
    primaryClass = "hep-ph",
    month = "9",
    year = "2026"
}

@article{Smirnov:2026aqk,
    author = "Smirnov, Juri and Griffith, Spencer and Beacom, John F.",
    title = "{Inelastic Signatures of Electroweak Dark Matter}",
    journal = "{arXiv e-prints}",
    eprint = "2609.04144",
    archivePrefix = "arXiv",
    primaryClass = "hep-ph",
    month = "9",
    year = "2026"
}

@article{deLima:2026shq,
    author = "de Lima, Carlos Henrique",
    title = "{Exothermic Dark Matter at LZ}",
    journal = "{arXiv e-prints}",
    eprint = "2609.05204",
    archivePrefix = "arXiv",
    primaryClass = "hep-ph",
    month = "9",
    year = "2026"
}

@article{Gu:2026vto,
    author = "Gu, Guanhua and Li, Lingfeng and Tang, Shao-Song and Xu, Yongheng",
    title = "{Inelastic from the Other Side: Xenon Excitation Signals in Light of the LZ High-Recoil Event}",
    journal = "{arXiv e-prints}",
    eprint = "2609.05291",
    archivePrefix = "arXiv",
    primaryClass = "hep-ph",
    month = "9",
    year = "2026"
}

@article{Dent:2026bji,
    author = "Dent, James B. and Newstead, Jayden L.",
    title = "{Exothermic and Endothermic Inelastic Dark Matter Interpretations at LZ: Sideband Constraints and Future Prospects}",
    journal = "{arXiv e-prints}",
    eprint = "2609.04673",
    archivePrefix = "arXiv",
    primaryClass = "hep-ph",
    month = "9",
    year = "2026"
}

@article{Lee:2026wof,
    author = "Lee, Hyun Min",
    title = "{Inelastic dark matter and baryon flavor symmetry in light of LUX-ZEPLIN (LZ) experiment}",
    journal = "{arXiv e-prints}",
    eprint = "2609.06171",
    archivePrefix = "arXiv",
    primaryClass = "hep-ph",
    month = "9",
    year = "2026"
}

@article{Das:2026uyy,
    author = "Das, Pritam and Karmakar, Biswajit and Mahapatra, Satyabrata and Paul, Partha Kumar",
    title = "{Inelastic Self-interacting Dark Matter and LUX-ZEPLIN 248 keV Event in a Dirac Modular Inverse Seesaw}",
    journal = "{arXiv e-prints}",
    eprint = "2609.06825",
    archivePrefix = "arXiv",
    primaryClass = "hep-ph",
    month = "9",
    year = "2026"
}

@article{DiMauro:2026dqp,
    author = "Di Mauro, Mattia and Shaikh, Halim",
    title = "{Solar Capture Tests of Inelastic Dark Matter after the LZ High-Recoil Event}",
    journal = "{arXiv e-prints}",
    eprint = "2609.06760",
    archivePrefix = "arXiv",
    primaryClass = "hep-ph",
    month = "9",
    year = "2026"
}

@article{Wang:2026ytg,
    author = "Wang, Lei and Xiao, Yang",
    title = "{The Inert Doublet Model of Dark Matter and the LUX-ZEPLIN High-Recoil Event}",
    journal = "{arXiv e-prints}",
    eprint = "2609.06571",
    archivePrefix = "arXiv",
    primaryClass = "hep-ph",
    month = "9",
    year = "2026"
}

@article{Bose:2026ndd,
    author = "Bose, Debajit and others",
    title = "{Not so good $\nu$s for Higgsino dark matter as LZ excess: stringent limits from Super-Kamiokande and IceCube}",
    journal = "{arXiv e-prints}",
    eprint = "2609.07807",
    archivePrefix = "arXiv",
    primaryClass = "hep-ph",
    month = "9",
    year = "2026"
}

@article{Bisal:2026khf,
    author = "Bisal, Subhadip and Cao, Junjie and Li, Fei",
    title = "{Higgsino Dark Matter Interpretation of the LZ High-Recoil Event in the GNMSSM with TeV-Scale Gauginos}",
    journal = "{arXiv e-prints}",
    eprint = "2609.07811",
    archivePrefix = "arXiv",
    primaryClass = "hep-ph",
    month = "9",
    year = "2026"
}

@article{Borah:2026zwf,
    author = "Borah, Debasish and Sahoo, Sujit Kumar and Sahu, Narendra and Sharma, Shashwat",
    title = "{Inelastic Singlet-Doublet Fermion Dark Matter in light of the 248 keV LZ event}",
    journal = "{arXiv e-prints}",
    eprint = "2609.07800",
    archivePrefix = "arXiv",
    primaryClass = "hep-ph",
    month = "9",
    year = "2026"
}

@article{Du:2026lpa,
    author = "Du, Xin-Yu and Huang, Wenjie and Xie, Keping",
    title = "{Pseudo-Dirac Inelastic Dark Matter in the Leptophobic $U(1)_B$ Model: Confronting the LUX-ZEPLIN High-Recoil Event with Collider Searches}",
    journal = "{arXiv e-prints}",
    eprint = "2609.07225",
    archivePrefix = "arXiv",
    primaryClass = "hep-ph",
    month = "9",
    year = "2026"
}

@article{Ahmed:2026qjg,
    author = "Ahmed, Waqas and Leontaris, George K.",
    title = "{A Dark-Dimension Origin of Geometric Inelastic Dark Matter: The LUX-ZEPLIN High-Recoil Event and Multi-Target Tests}",
    journal = "{arXiv e-prints}",
    eprint = "2609.07138",
    archivePrefix = "arXiv",
    primaryClass = "hep-ph",
    month = "9",
    year = "2026"
}

@article{Okada:2026eol,
    author = "Okada, Nobuchika and Seto, Osamu",
    title = "{Inelastic $B-L$ scalar dark matter and the LUX-ZEPLIN event}",
    journal = "{arXiv e-prints}",
    eprint = "2609.06909",
    archivePrefix = "arXiv",
    primaryClass = "hep-ph",
    reportNumber = "EPHOU-26-011",
    month = "9",
    year = "2026"
}

@article{Langhoff:2026ujr,
    author = "Langhoff, Kevin",
    title = "{Heavy Higgsino Interpretation of the LZ Event}",
    journal = "{arXiv e-prints}",
    eprint = "2609.09385",
    archivePrefix = "arXiv",
    primaryClass = "hep-ph",
    month = "9",
    year = "2026"
}

@article{Asadi:2026iot,
    author = "Asadi, Pouya and Batz, Austin and Fox, Patrick J. and Homiller, Samuel D. and Kribs, Graham D.",
    title = "{For Whom the Xenon Recoils: Magnetic Inelastic Dark Baryons}",
    journal = "{arXiv e-prints}",
    eprint = "2609.09107",
    archivePrefix = "arXiv",
    primaryClass = "hep-ph",
    reportNumber = "PITT-PACC-2614, FERMILAB-PUB-26-0660-T",
    month = "9",
    year = "2026"
}

@article{Lee:2026jxl,
    author = "Lee, Seung J. and Youn, Taewook",
    title = "{Mixing-suppressed inelastic dark matter: a minimal model for the LZ 248 keV event}",
    journal = "{arXiv e-prints}",
    eprint = "2609.09138",
    archivePrefix = "arXiv",
    primaryClass = "hep-ph",
    month = "9",
    year = "2026"
}

@article{Cheung:2026byg,
    author = "Cheung, Kingman and Kang, Sin Kyu and Kumar, Ranjeet",
    title = "{From LUX-ZEPLIN to Colliders: Probing Higgsino Dark Matter}",
    journal = "{arXiv e-prints}",
    eprint = "2609.08712",
    archivePrefix = "arXiv",
    primaryClass = "hep-ph",
    month = "9",
    year = "2026"
}

@article{Yuan:2026djt,
    author = "Yuan, Guan-Wen and Zhang, Bo and Cao, Wen-Yu and Feng, Lei and Yang, Ruizhi",
    title = "{ALP-mediated inelastic dark matter and the LUX-ZEPLIN high-recoil candidate event LZ230616}",
    journal = "{arXiv e-prints}",
    eprint = "2609.08893",
    archivePrefix = "arXiv",
    primaryClass = "hep-ph",
    month = "9",
    year = "2026"
}

@article{Baer:2026fpy,
    author = "Baer, Howard and Barger, Vernon",
    title = "{Exothermic dark matter and the 248 keV nuclear recoil in LUX-ZEPLIN}",
    journal = "{arXiv e-prints}",
    eprint = "2609.06153",
    archivePrefix = "arXiv",
    primaryClass = "hep-ph",
    month = "9",
    year = "2026"
}

@article{Kumar:2026lgi,
    author = "Kumar, Ranjeet and Prajapati, Hemant Kumar",
    title = "{Generalized Chiral $U(1)_{B-L}$ with Inelastic Scalar Dark Matter for the LZ 248 keV Event}",
    journal = "{arXiv e-prints}",
    eprint = "2609.10827",
    archivePrefix = "arXiv",
    primaryClass = "hep-ph",
    month = "9",
    year = "2026"
}

@article{Qi:2026vyp,
    author = "Qi, XinXin and Sun, Hao",
    title = "{Nonthermal Solar Stalling of an Inelastic Scalar Signal in Xenon}",
    journal = "{arXiv e-prints}",
    eprint = "2609.10636",
    archivePrefix = "arXiv",
    primaryClass = "hep-ph",
    month = "9",
    year = "2026"
}

@article{Fan:2026hzw,
    author = "Fan, Zi-Tong and He, Hong-Jian and Wang, Yu-Chen and Zhao, Yue",
    title = "{Inelastic Dark Matter and High-Energy Recoil Signatures in LZ}",
    journal = "{arXiv e-prints}",
    eprint = "2609.10491",
    archivePrefix = "arXiv",
    primaryClass = "hep-ph",
    month = "9",
    year = "2026"
}

@article{Chatterjee:2026scv,
    author = "Chatterjee, Arindam and Das, Debottam and Pasha, Syed Adil and Pukhov, Alexander and Puri, Rahul",
    title = "{Radiative Corrections to the Direct Detection of Inelastic Scattering of Higgsino-like Neutralino Dark Matter}",
    journal = "{arXiv e-prints}",
    eprint = "2609.09830",
    archivePrefix = "arXiv",
    primaryClass = "hep-ph",
    month = "9",
    year = "2026"
}

@article{Nguyen:2026lui,
    author = "Nguyen, Thong T. Q. and Linden, Tim and Hooper, Dan",
    title = "{Solar Neutrino Constraints on Inelastic Dark Matter Scattering in Light of Recent LUX-ZEPLIN Observations}",
    journal = "{arXiv e-prints}",
    eprint = "2609.11833",
    archivePrefix = "arXiv",
    primaryClass = "hep-ph",
    month = "9",
    year = "2026"
}

@article{He:2026idw,
    author = "He, Xiao-Gang and Hong, Xuan and Jeesun, Sk",
    title = "{Hadrophilic inelastic freeze-in dark matter in $q_1-q_2$ gauge extension and the high energy LZ event}",
    journal = "{arXiv e-prints}",
    eprint = "2609.15714",
    archivePrefix = "arXiv",
    primaryClass = "hep-ph",
    month = "9",
    year = "2026"
}

@article{Ahmed:2026kan,
    author = "Ahmed, Waqas and Ahmad, Ammara and Rehman, Mansoor Ur",
    title = "{Xenon Isotope Filtering at the Kinematic Edge of Inelastic Dark Matter}",
    journal = "{arXiv e-prints}",
    eprint = "2609.15634",
    archivePrefix = "arXiv",
    primaryClass = "hep-ph",
    month = "9",
    year = "2026"
}

@article{Lian:2026hpm,
    author = "Lian, Jingwei and Yang, Jin Min",
    title = "{Explain the LZ High-Energy Recoil Event with Inelastic Sneutrino Dark Matter in Supersymmetry}",
    journal = "{arXiv e-prints}",
    eprint = "2609.15742",
    archivePrefix = "arXiv",
    primaryClass = "hep-ph",
    month = "9",
    year = "2026"
}

@article{Das:2026buc,
    author = "Das, Arindam and Nomura, Takaaki",
    title = "{Effect of inelastic scalar dark matter in hidden $U(1)$ scenario after the LZ nuclear recoil}",
    journal = "{arXiv e-prints}",
    eprint = "2609.15600",
    archivePrefix = "arXiv",
    primaryClass = "hep-ph",
    month = "9",
    year = "2026"
}

@article{Ghosh:2026txe,
    author = "Ghosh, Aditya and Chavez, Ilumi and Kelso, Chris",
    title = "{Confronting the Higgsino Interpretation of the LZ Event with Astrophysical Uncertainties and the Solar Capture Constraints}",
    journal = "{arXiv e-prints}",
    eprint = "2609.15321",
    archivePrefix = "arXiv",
    primaryClass = "hep-ph",
    month = "9",
    year = "2026"
}

@article{Borah:2026ris,
    author = "Borah, Pankaj and Mahapatra, Satyabrata and Nath, Newton and Paul, Partha Kumar",
    title = "{Inelastic Dark Matter at LZ from Radiative Dirac Neutrino Mass Paradigm}",
    journal = "{arXiv e-prints}",
    eprint = "2609.15027",
    archivePrefix = "arXiv",
    primaryClass = "hep-ph",
    month = "9",
    year = "2026"
}

@article{Ge:2026xax,
    author = "Ge, Shao-Feng and Titov, Oleg and Wang, Yakun",
    title = "{Dark Matter Inelastic Scattering with Nuclei for Direct Detection}",
    journal = "{arXiv e-prints}",
    eprint = "2609.16529",
    archivePrefix = "arXiv",
    primaryClass = "hep-ph",
    month = "9",
    year = "2026"
}

@misc{rodd2026,
      title={Confronting the Higgsino Interpretation of the LZ Event with the High-Energy Sideband}, 
      author={Nicholas L. Rodd and Benjamin R. Safdi and Tracy R. Slatyer and Weishuang Linda Xu},
      year={2026},
      eprint={2609.04175},
      archivePrefix={arXiv},
      primaryClass={hep-ph},
      url={https://arxiv.org/abs/2609.04175}, 
}

@misc{jeesun2026,
      title={Atmospheric neutrino up-scattering explanation of LZ 2026 excess}, 
      author={Sk Jeesun and Anirban Majumdar},
      year={2026},
      eprint={2609.04185},
      archivePrefix={arXiv},
      primaryClass={hep-ph},
      url={https://arxiv.org/abs/2609.04185}, 
}

@misc{fan2026,
      title={Higgsino Above the Sea of Fog}, 
      author={JiJi Fan and Matthew Reece},
      year={2026},
      eprint={2609.01504},
      archivePrefix={arXiv},
      primaryClass={hep-ph},
      url={https://arxiv.org/abs/2609.01504}, 
}

@misc{bandyopadhyay2026,
      title={LZ nuclear recoil event from inelastic singlet-doublet scalar dark matter}, 
      author={Disha Bandyopadhyay and Debasish Borah and Pankaj Borah},
      year={2026},
      eprint={2609.07451},
      archivePrefix={arXiv},
      primaryClass={hep-ph},
      url={https://arxiv.org/abs/2609.07451}, 
}

@article{Nagata:2026pbj,
    author = "Nagata, Natsumi and Yanagida, Tsutomu T.",
    title = "{Asymmetric Inelastic Dark Matter and the LUX-ZEPLIN event}",
    journal = "{arXiv e-prints}",
    eprint = "2609.18564",
    archivePrefix = "arXiv",
    primaryClass = "hep-ph",
    month = "9",
    year = "2026"
}

@article{Xing:2026civ,
    author = "Xing, Chuan-Yang",
    title = "{Galactic Endothermic Production and Exothermic Detection of,  Excited Dark Matter: Implications for LUX-ZEPLIN}",
    journal = "{arXiv e-prints}",
    eprint = "2609.17935",
    archivePrefix = "arXiv",
    primaryClass = "hep-ph",
    month = "9",
    year = "2026"
}

@article{An:2026pkc,
    author = "An, Haipeng and Gao, Fei and Liu, Jia and Liu, Minghao and Xu, Changlong",
    title = "{Cosmological Constrained Axion-Portal Inelastic Dark Matter for the LZ Event}",
    journal = "{arXiv e-prints}",
    eprint = "2609.17412",
    archivePrefix = "arXiv",
    primaryClass = "hep-ph",
    month = "9",
    year = "2026"
}

@misc{yin2026,
      title={A PQ-Symmetric High-Scale SUSY Interpretation of the LZ High-Energy Recoil}, 
      author={Wen Yin},
      year={2026},
      eprint={2609.01892},
      archivePrefix={arXiv},
      primaryClass={hep-ph},
      url={https://arxiv.org/abs/2609.01892}, 
}

@article{OHare:2026nqi,
    author = "O'Hare, Ciaran A. J.",
    title = "{The high-velocity dark matter halo of the Milky Way in light of the LZ 248 keV event}",
    journal = "{arXiv e-prints}",
    eprint = "2609.21444",
    archivePrefix = "arXiv",
    primaryClass = "hep-ph",
    month = "9",
    year = "2026"
}

@article{Liang:2026coz,
    author = "Liang, Jin-Han and Liu, Zuowei and Tran, Van Que and Xu, Yongheng",
    title = "{LZ Nuclear-Recoil Excess from Boosted Light Magnetic Dipole-dipole Dark Matter}",
    journal = "{arXiv e-prints}",
    eprint = "2609.06756",
    archivePrefix = "arXiv",
    primaryClass = "hep-ph",
    month = "9",
    year = "2026"
}

@article{Kannike:2026qyl,
    author = "Kannike, Kristjan and Raidal, Martti and Strumia, Alessandro",
    title = "{Boosted dark particles and the LZ nuclear recoil event}",
    journal = "{arXiv e-prints}",
    eprint = "2609.07742",
    archivePrefix = "arXiv",
    primaryClass = "hep-ph",
    month = "9",
    year = "2026"
}

@article{Alhazmi:2026efz,
    author = "Alhazmi, Haider and Kim, Doojin and Kong, Kyoungchul and Park, Jong-Chul and Shin, Seodong",
    title = "{High-Energy Nuclear Recoils from Boosted Dark Matter for the LZ 248-keV Event: Beyond the Halo-Dependent High-Velocity Tail}",
    journal = "{arXiv e-prints}",
    eprint = "2609.06890",
    archivePrefix = "arXiv",
    primaryClass = "hep-ph",
    month = "9",
    year = "2026"
}

@article{Heikinheimo:2026kwp,
    author = "Heikinheimo, Matti and Zimmermann, Niklas",
    title = "{Cosmic ray boosted dark matter with momentum dependent interactions can explain the LZ 248 keV event}",
    journal = "{arXiv e-prints}",
    eprint = "2609.11600",
    archivePrefix = "arXiv",
    primaryClass = "hep-ph",
    month = "9",
    year = "2026"
}

@article{Mahapatra:2026glu,
    author = "Mahapatra, Satyabrata and Paul, Partha Kumar",
    title = "{Boosted or Inelastic? Discriminating Interpretations of the LZ 248 keV Event}",
    journal = "{arXiv e-prints}",
    eprint = "2609.14799",
    archivePrefix = "arXiv",
    primaryClass = "hep-ph",
    month = "9",
    year = "2026"
}

@article{Unwin:2026rdp,
    author = "Unwin, James",
    title = "{Axion Portal Dark Matter and the LUX-ZEPLIN High-Recoil Event}",
    journal = "{arXiv e-prints}",
    eprint = "2609.04186",
    archivePrefix = "arXiv",
    primaryClass = "hep-ph",
    month = "9",
    year = "2026"
}

@article{Kotlarski:2026pep,
    author = "Kotlarski, Wojciech and Kowalska, Kamila and Sessolo, Enrico Maria",
    title = "{GUT-induced FCC signatures of the LUX-ZEPLIN event}",
    journal = "{arXiv e-prints}",
    eprint = "2609.06750",
    archivePrefix = "arXiv",
    primaryClass = "hep-ph",
    month = "9",
    year = "2026"
}

@article{Yang:2026wpb,
    author = "Yang, Meiwen and Wu, Quan-feng and Tsai, Yue-Lin Sming and Fan, Yi-Zhong",
    title = "{Multi-Messenger and Paleo-Detector Probes of the LZ Dark Matter Signal}",
    journal = "{arXiv e-prints}",
    eprint = "2609.06640",
    archivePrefix = "arXiv",
    primaryClass = "hep-ph",
    month = "9",
    year = "2026"
}

@article{Khan:2026nwp,
    author = "Khan, Imtiaz and Capozziello, Salvatore and Mustafa, G. and Atamurotov, Farruh and Abdujabbarov, Ahmadjon and Yuan, Chengxun",
    title = "{Nuclear interference versus dark sector excitation in the 248 keV LUX-ZEPLIN recoil candidate}",
    journal = "{arXiv e-prints}",
    eprint = "2609.09230",
    archivePrefix = "arXiv",
    primaryClass = "hep-ph",
    month = "9",
    year = "2026"
}

@article{Lee:2026xxh,
    author = "Lee, Vincent S. H. and Randall, Lisa",
    title = "{A Warped Extra Dimensional Candidate for the LZ 248 keV Event}",
    journal = "{arXiv e-prints}",
    eprint = "2609.09136",
    archivePrefix = "arXiv",
    primaryClass = "hep-ph",
    reportNumber = "N3AS-26-020",
    month = "9",
    year = "2026"
}

@article{Aghaie:2026vsu,
    author = "Aghaie, Mohammad and Strumia, Alessandro",
    title = "{Neutron disappearance and the LZ nuclear recoil event}",
    journal = "{arXiv e-prints}",
    eprint = "2609.09037",
    archivePrefix = "arXiv",
    primaryClass = "hep-ph",
    month = "9",
    year = "2026"
}

@article{Zhu:2026dag,
    author = "Zhu, Pengxuan and Dalla Valle Garcia, Giovani and Wang, Xuan-Gong and Thomas, Anthony W. and White, Martin J.",
    title = "{Endothermic dark matter with a light dark photon and the LUX--ZEPLIN high-energy nuclear-recoil candidate}",
    journal = "{arXiv e-prints}",
    eprint = "2609.09015",
    archivePrefix = "arXiv",
    primaryClass = "hep-ph",
    month = "9",
    year = "2026"
}

@article{Elahi:2026vlm,
    author = "Elahi, Fatemeh and Schwaller, Pedro",
    title = "{A Vector-Like Lepton Interpretation of the High-Energy Nuclear Recoil Candidate in LUX-ZEPLIN}",
    journal = "{arXiv e-prints}",
    eprint = "2609.08993",
    archivePrefix = "arXiv",
    primaryClass = "hep-ph",
    reportNumber = "MITP-26-043",
    month = "9",
    year = "2026"
}

@article{Bamwidhi:2026vdu,
    author = "Bamwidhi, Isaac and Belfkir, Mohamed and Loualidi, Mohamed Amin and Nasri, Salah",
    title = "{Neutrino mass, scalar dark matter, and collider signatures in a radiative doublet-triplet model}",
    journal = "{arXiv e-prints}",
    eprint = "2609.08808",
    archivePrefix = "arXiv",
    primaryClass = "hep-ph",
    month = "9",
    year = "2026"
}

@article{Egorov:2026dpr,
    author = "Egorov, Andrei E.",
    title = "{Detection prospects for heavy WIMP dark matter around M31* in microwave band}",
    journal = "{arXiv e-prints}",
    eprint = "2609.08771",
    archivePrefix = "arXiv",
    primaryClass = "astro-ph.GA",
    month = "9",
    year = "2026"
}

@article{He:2026hqz,
    author = "He, Yuxuan",
    title = "{Transition magnetic-dipole dark matter and the LZ230616 high-recoil candidate}",
    journal = "{arXiv e-prints}",
    eprint = "2609.10453",
    archivePrefix = "arXiv",
    primaryClass = "hep-ph",
    month = "9",
    year = "2026"
}

@article{Chattaraj:2026fxn,
    author = "Chattaraj, Ayan and Majumdar, Anirban and Papoulias, Dimitrios K. and Srivastava, Rahul",
    title = "{Can Elastic Neutrino Scattering Account for the LZ230616 Event?}",
    journal = "{arXiv e-prints}",
    eprint = "2609.10504",
    archivePrefix = "arXiv",
    primaryClass = "hep-ph",
    month = "9",
    year = "2026"
}

@article{Lee:2026zbr,
    author = "Lee, Junseok and Takahashi, Fuminobu and Tsai, Yu-Dai",
    title = "{Nuclear Recoils from Invisible Neutron-Pair Annihilation and the LZ event}",
    journal = "{arXiv e-prints}",
    eprint = "2609.12045",
    archivePrefix = "arXiv",
    primaryClass = "hep-ph",
    month = "9",
    year = "2026"
}

@article{Okada:2026upm,
    author = "Okada, Hiroshi and Shigekami, Yoshihiro and Wu, Jia-Jun",
    title = "{Can a minimal radiative seesaw explain the LZ 248 keV event?}",
    journal = "{arXiv e-prints}",
    eprint = "2609.13038",
    archivePrefix = "arXiv",
    primaryClass = "hep-ph",
    month = "9",
    year = "2026"
}

@article{DiMauro:2026ymt,
    author = "Di Mauro, Mattia",
    title = "{Testing Higgs-Coupled Minimal Dark Matter with Solar Neutrinos after the LZ High-Recoil Event}",
    journal = "{arXiv e-prints}",
    eprint = "2609.19174",
    archivePrefix = "arXiv",
    primaryClass = "hep-ph",
    month = "9",
    year = "2026"
}

@article{Uttayarat:2026isp,
    author = "Uttayarat, P. and Julio, J. and Primulando, R.",
    title = "{DM induced neutron disappearance as the origin of the LZ nuclear recoil event}",
    journal = "{arXiv e-prints}",
    eprint = "2609.15933",
    archivePrefix = "arXiv",
    primaryClass = "hep-ph",
    month = "9",
    year = "2026"
}

@article{Palmisano:2026kuj,
    author = "Palmisano, Stefano and Tammaro, Michele and Tesi, Andrea",
    title = "{Inferring dark matter masses and interactions from high recoil energy events in LUX-ZEPLIN}",
    journal = "{arXiv e-prints}",
    eprint = "2609.15985",
    archivePrefix = "arXiv",
    primaryClass = "hep-ph",
    month = "9",
    year = "2026"
}

@article{Barman:2026omh,
    author = "Barman, Basabendu",
    title = "{Did LZ see modified gravity?}",
    journal = "{arXiv e-prints}",
    eprint = "2609.15118",
    archivePrefix = "arXiv",
    primaryClass = "hep-ph",
    month = "9",
    year = "2026"
}

@article{Arcadi:2026kev,
    author = "Arcadi, Giorgio and di Mauro, Mattia and Djouadi, Abdelhak and Queiroz, Farinaldo",
    title = "{A possible interpretation of the LUX-ZEPLIN recoil event in the 2HD+a scenario}",
    journal = "{arXiv e-prints}",
    eprint = "2609.17196",
    archivePrefix = "arXiv",
    primaryClass = "hep-ph",
    month = "9",
    year = "2026"
}

@article{Lueiza-Colipi:2026gij,
    author = "Lueiza-Colip{\'\i}, Andr{\'e}s and Dimakis, Nikolaos and Leon, Genly and Paliathanasis, Andronikos",
    title = "{Symmetry-Driven $k$-Essence Cosmological Dynamics}",
    journal = "{arXiv e-prints}",
    eprint = "2609.18790",
    archivePrefix = "arXiv",
    primaryClass = "gr-qc",
    month = "9",
    year = "2026"
}

@misc{dimauro2026,
      title={Dark Matter at the Kinematic Edge: Interpreting the 248 keV LZ Nuclear-Recoil Candidate}, 
      author={Mattia Di Mauro},
      year={2026},
      eprint={2609.02608},
      archivePrefix={arXiv},
      primaryClass={hep-ph},
      url={https://arxiv.org/abs/2609.02608}, 
}

@misc{mccabe2026,
      title={Seasonal dark matter from the LUX-ZEPLIN high-energy event}, 
      author={Christopher McCabe},
      year={2026},
      eprint={2609.04181},
      archivePrefix={arXiv},
      primaryClass={hep-ph},
      url={https://arxiv.org/abs/2609.04181}, 
}

@article{Bose:2026szs,
    author = "Bose, Debajit and Saha, Akash Kumar and Raj, Nirmal and Maity, Tarak Nath and Laha, Ranjan",
    title = "{LUX-ZEPLIN's Stairway to Hea$\nu$en: limits on elastic scatters of dark matter from solar capture}",
    journal = "{arXiv e-prints}",
    eprint = "2609.21823",
    archivePrefix = "arXiv",
    primaryClass = "hep-ph",
    month = "9",
    year = "2026"
}

@article{Brdar:2026ukx,
    author = "Brdar, Vedran and Chattopadhyay, Dibya S.",
    title = "{A Narrow Neutrino Window for the LZ Event}",
    journal = "{arXiv e-prints}",
    eprint = "2609.30255",
    archivePrefix = "arXiv",
    primaryClass = "hep-ph",
    month = "9",
    year = "2026"
}

@article{Sayan2019,
  author = {{Mandal}, Sayan and {Majumdar}, Subhabrata and {Rentala}, Vikram and {Basu Thakur}, Ritoban},
  title = "{Observationally inferred dark matter phase-space distribution and direct detection experiments}",
  journal = {Physical Review D},
  volume = {100},
  number = {2},
  pages = {023002},
  year = {2019},
  month = {Jul},
  doi = {10.1103/PhysRevD.100.023002},
  archivePrefix = {arXiv},
  eprint = {1806.05629},
  primaryClass = {astro-ph.GA},
  url = {https://ui.adsabs.harvard.edu/abs/2019PhRvD.100b3002M}
}
\clearpage
\appendix

\section{Methodology}
\label{sec:analysis-notes}

This section provides supplementary details on the detector efficiency and the profile-likelihood evaluation used in the main text. The complete list of analysis inputs is collected in Table~\ref{tab:notes-inputs}.

\subsection{Detector efficiency and search regions}
\label{sec:notes-data}

As discussed in the main text, LZ defines their search region of interest (ROI) as $5.4 < E_R < 269.9\keV$, where the efficiency averages $96\%$ between $14$ and $250\keV$. We model this efficiency $\epsilon(E_R)$ as a flat plateau with smooth (error-function) edges at the two 50\% points:
\begin{align}
  \epsilon(E_R) = 0.96
  &\times \tfrac{1}{2}\!\left[1+{\rm erf}\!\left(\tfrac{E_R-5.4\keV}{\sqrt2\,\sigma_{\rm lo}}\right)\right] \nonumber\\
  &\times \tfrac{1}{2}\!\left[1+{\rm erf}\!\left(\tfrac{269.9\keV-E_R}{\sqrt2\,\sigma_{\rm hi}}\right)\right],
  \label{eq:eff}
\end{align}
with widths $\sigma_{\rm lo} = 3\keV$ and $\sigma_{\rm hi} = 6\keV$. Because LZ gives no efficiency above the ROI, we set $\epsilon = 0$ above $270\keV$ (a ``hard cut'') when evaluating the search data in regions S and W. For the high-energy sideband H ($350$--$650\keV$), we use a flat efficiency of $0.96$.

\begin{table*}[t]
  \caption{Analysis Inputs}
  \label{tab:notes-inputs}
  \centering
  \scriptsize
  \begin{ruledtabular}
  \begin{tabular}{ll}
    quantity & value \\
    \hline
    \multicolumn{2}{l}{\textit{Detector and data}} \\
    event energy $E_1$ & $248\keV$ \\
    energy uncertainty $\sigma_E$ & $\sqrt{23^2+23^2}=32.5\keV$ \\
    exposure $MT$ & 2.84\,t\,yr $=1.037\times10^6$\,kg\,d \\
    ROI (50\% efficiency points) & $5.4$--$269.9\keV$ \\
    efficiency plateau & 0.96 \\
    edge widths $\sigma_{\rm lo},\sigma_{\rm hi}$ & $3,\ 6\keV$ \\
    background near event & $0.0106\pm0.0008$ \\
    high-energy sideband & $350$--$650\keV$, 0 events \\
    \hline
    \multicolumn{2}{l}{\textit{Counting regions}} \\
    S (low sideband) & $5.4$--$215.5\keV$, 0 events \\
    W (event window) & $215.5$--$280.5\keV$, 1 event \\
    H (high sideband) & $350$--$650\keV$, 0 events \\
    \hline
    \multicolumn{2}{l}{\textit{Nuclear physics}} \\
    isotopes & $^{124,126,128\text{--}132,134,136}$Xe, natural abundances \\
    nuclear response & shell-model $W_M^{00}(q)$ \\
    \hline
    \multicolumn{2}{l}{\textit{Halo: SHM}} \\
    $v_0$, $\vesc$ & $238,\ 544\,\rm km\,s^{-1}$ \\
    solar peculiar motion & $(9,12,7)\,\rm km\,s^{-1}$ \\
    $v_{\rm lab}$ & $250.3\,\rm km\,s^{-1}$ \\
    $\rho_\chi$ & $0.3\GeV\,{\rm cm^{-3}}$ \\
    \hline
    \multicolumn{2}{l}{\textit{Halo: Einasto VDF (best-fit mass model)}} \\
    $\vesc$ & $521.8\,\rm km\,s^{-1}$ \\
    $v_{\rm lab}$ & $246.7\,\rm km\,s^{-1}$ \\
    $\rho_\chi$ & $0.3$ and $0.567\GeV\,{\rm cm^{-3}}$ \\
    \hline
    \multicolumn{2}{l}{\textit{Analysis choices}} \\
    endothermic edge rule & $v_{\rm cut}-v_* \ge 20\,\rm km\,s^{-1}$ \\
    $\Delta(-2\ln\mathcal{L})$, 2 params & 2.30 (68\%), 5.99 (95\%) \\
    $\Delta(-2\ln\mathcal{L})$, 1 param & 1.00 (68\%), 3.84 (95\%) \\
  \end{tabular}
  \end{ruledtabular}
\end{table*}

\subsection{Profile Likelihood}
\label{sec:notes-likelihood}

Without the high-energy sideband H, the likelihood cannot distinguish a good model from the one that puts almost all of its events above $270\keV$, where the search data have zero efficiency. For example, at $m_\chi = 100\GeV$, the best exothermic fit without H predicts $1{,}479$ (SHM) or $2{,}467$ (Einasto) events in $350$--$650\keV$ for every event in W. Since LZ saw zero events there, including H effectively eliminates these pathological solutions.

Using the likelihood $\mathcal{L}$ defined in Eq.~\eqref{eq:likelihood}, every expected count $N_X$ is proportional to $\sigma_p\rho_\chi$. We write $N_{\rm W} = sA$, $N_{\rm S} = sB$, $N_{\rm H} = sC$, where $s \equiv \sigma_p[{\rm cm^2}]\,\rho_\chi/(0.3\GeV\,{\rm cm^{-3}})$ and $A$, $B$, $C$ are the expected counts per unit $s$, which depend only on $(m_\chi,\delta)$ and the VDF. The negative log-likelihood then becomes:
\begin{equation}
  -2\ln\mathcal{L} = 2s(A+B+C) - 2\ln(sA).
  \label{eq:m2lnl}
\end{equation}
Three consequences follow from this formulation:
\begin{enumerate}
  \item Equation~\eqref{eq:m2lnl} is minimised at $\hat s = 1/(A+B+C)$, i.e., when the three regions together contain exactly one expected event. At this minimum:
  \begin{equation}
    -2\ln\mathcal{L}_{\rm prof} = 2 - 2\ln F, \quad
    F \equiv \frac{N_{\rm W}}{N_{\rm S}+N_{\rm W}+N_{\rm H}},
  \end{equation}
  where $F$ is the \emph{fraction} of expected events that land in the event window. The best model for a given $(m_\chi,\delta)$ is simply the one with the largest $F$.
  \item $F$ does not depend on $\rho_\chi$, so the best-fit $\delta$, all $(m_\chi,\delta)$ contours, and the $\vesc$--$\delta$ degeneracy are the same for $\rho_\chi = 0.3$ and $0.567\GeV\,{\rm cm^{-3}}$. Only $\hat\sigma_p$ changes, scaling by exactly $0.3/\rho_\chi$ (a factor of $1/1.89$).
  \item Since $F\le1$, it implies $-2\ln\mathcal{L}\ge2$. The floor is reached only when \emph{all} expected events fall in W, which occurs for exothermic DM lighter than about $10\GeV$, where the spectrum is narrow enough to fit entirely inside W.
\end{enumerate}

We quote regions where $-2\ln\mathcal{L}$ lies less than $\Delta$ above its minimum, with $\Delta = 2.30$ (68\%) and $5.99$ (95\%) for two parameters and $1.00$ and $3.84$ for one parameter. In terms of $F$, a point is inside the 68\% (95\%) region if $F \ge 0.32\,F_{\rm max}$ ($0.05\,F_{\rm max}$). Finally, regarding the evaluation of the endothermic parameter space: as $\delta\to\delta_{\rm edge}$, the spectrum shrinks to a narrow sliver at $E_*$ and the rate approaches zero. In our numerical analysis, we scan the variable $s \equiv v_*/v_{\rm cut}$ and set $\delta = \delta_{\rm edge}s^2$.

\subsection{Results}

\begin{figure}[t]
  \centering
  \includegraphics[width=\columnwidth]{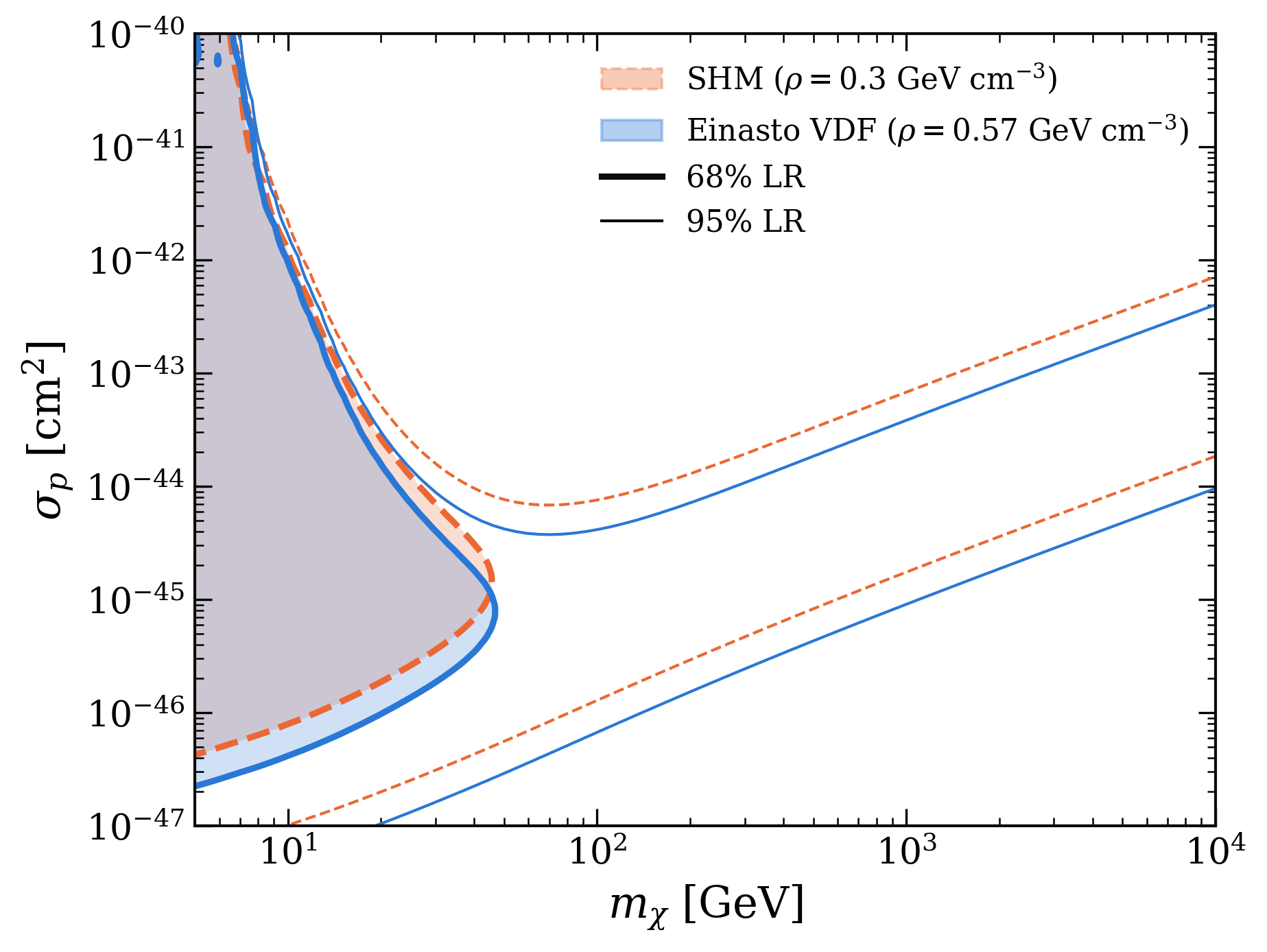}
  \caption{Allowed regions in $(m_\chi,\sigma_p)$ for exothermic scattering,
  with $\delta$ profiled, for the SHM and the Einasto VDF. Contours are 68\% and 95\%.}
  \label{fig:regions_exo}
\end{figure}

At each point of this plane $\delta$ is chosen to give the best fit (``profiled''), and contours are drawn relative to each halo's own best fit (Figs.~\ref{fig:regions_exo} and~\ref{fig:regions_endo}).

For the exothermic case, at the same density, the inferred $\sigma_p$ changes by a factor $1.02$--$1.11$ for $m_\chi \ge 30\GeV$, and the best-fit $\delta$ by at most $2.4\%$. The two 68\% regions nearly coincide. Using the fitted Einasto density moves the whole region down by the factor $1.89$.

\begin{figure}[t]
  \centering
  \includegraphics[width=\columnwidth]{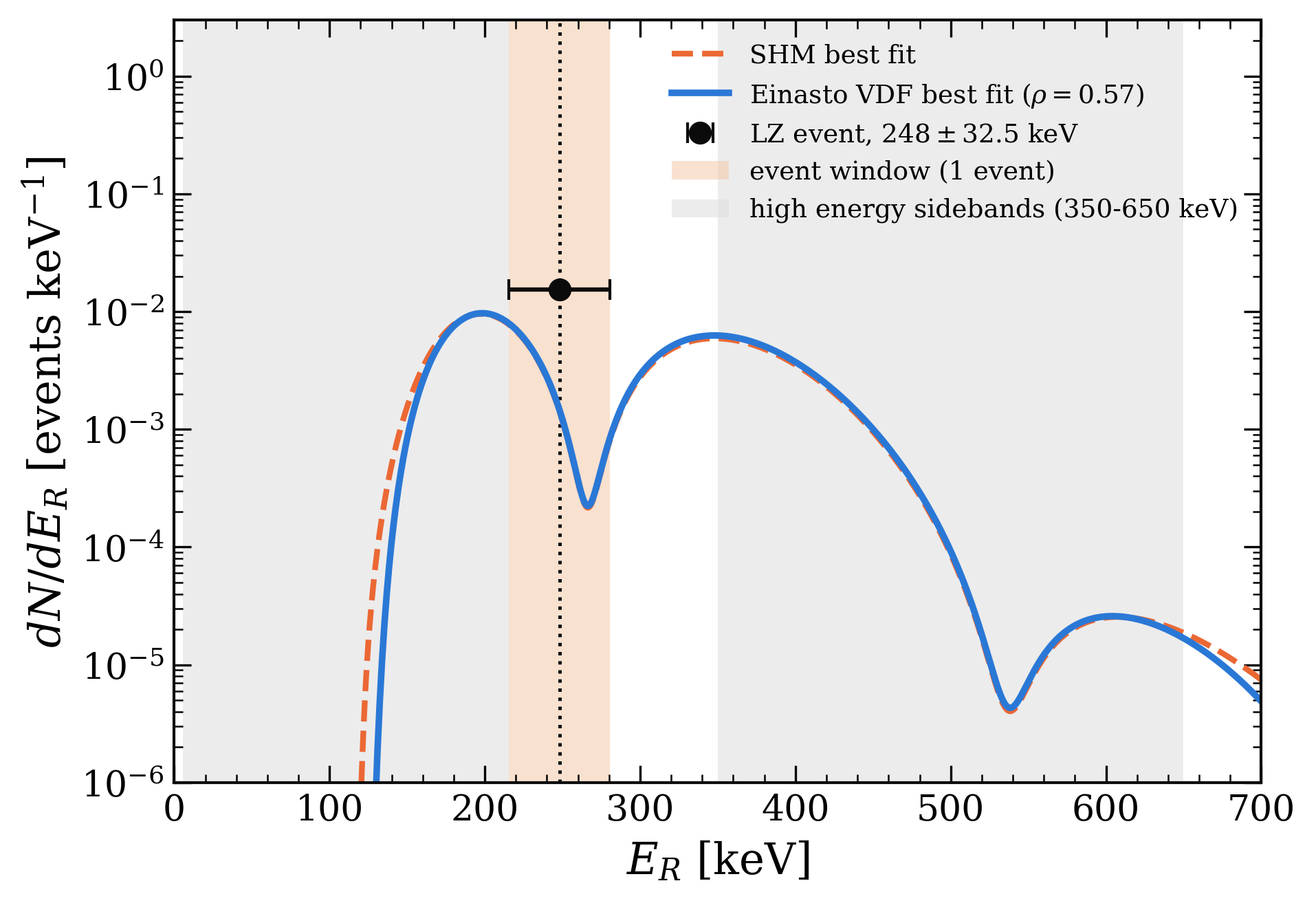}
  \caption{Predicted recoil spectra for the exothermic best fits at
  $m_\chi = 100\GeV$ (SHM and Einasto VDF). Shaded bands mark the regions S,
  W and H; the black point is the LZ event with its $\pm1\sigma$ energy
  uncertainty.}
  \label{fig:notes-spec-exo}
\end{figure}

\begin{figure}[t]
  \centering
  \includegraphics[width=\columnwidth]{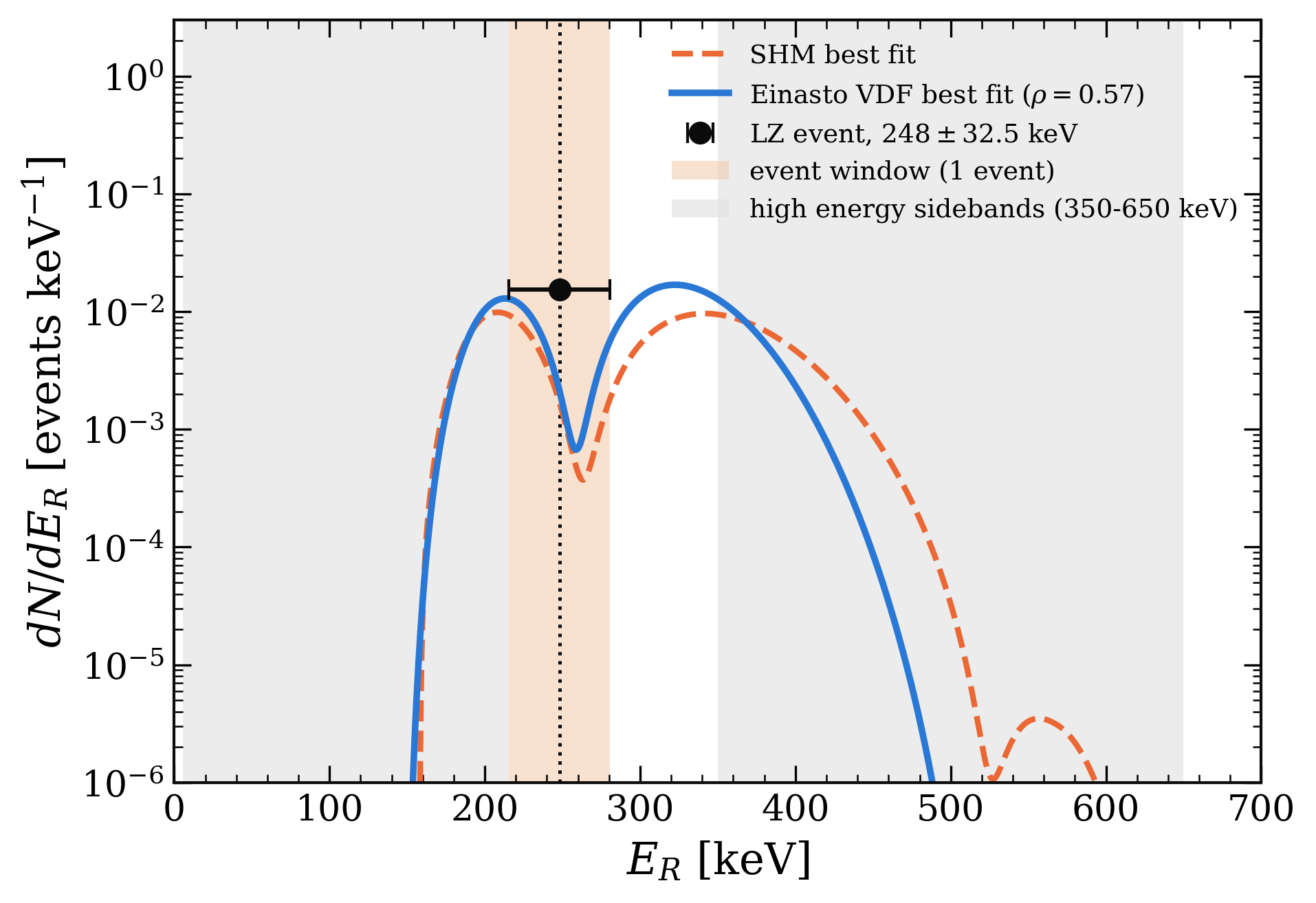}
  \caption{As Fig.~\ref{fig:notes-spec-exo}, for the endothermic best fits at
  $m_\chi = 1\,{\rm TeV}$.}
  \label{fig:notes-spec-endo}
\end{figure}

Figures~\ref{fig:notes-spec-exo} and~\ref{fig:notes-spec-endo} show the predicted recoil
spectrum, in events per keV, for the best fit of each halo at
$m_\chi = 100\GeV$ (exothermic) and $1\,{\rm TeV}$ (endothermic).
Table~\ref{tab:notes-spectra} lists the parameters and the expected counts in
each region. For plotting, the curves use the efficiency of
Eq.~\eqref{eq:eff} \emph{without} the $270\keV$ hard cut, so that the whole
spectrum is visible. The LZ event is drawn as a black point at $248\keV$ with
horizontal bar $\pm\sigma_E$. Its height, $1/(2\sigma_E) = 0.0154$ events per keV,
is one event spread evenly over W, so a model that put exactly one event in W
would pass through it on average.

\begin{table}[t]
  \caption{Benchmark fits used for the spectrum figures. $N_{\rm S}$, $N_{\rm W}$
  and $N_{\rm H}$ are the expected counts at $\hat\sigma_p$; they add up to one
  by construction. The Einasto fits use the fitted density
  $\rho_\chi = 0.567\GeV\,{\rm cm^{-3}}$.}
  \label{tab:notes-spectra}
  \centering
  \scriptsize
  \begin{ruledtabular}
  \setlength{\tabcolsep}{3pt}
  \begin{tabular}{llccc}
    & model & $\delta$ [keV] & $\hat\sigma_p$ [cm$^2$] & $N_{\rm S}/N_{\rm W}/N_{\rm H}$ \\
    \hline
    \multirow{2}{*}{exo, 100\,GeV}
      & SHM        & $-722$ & $1.61\times10^{-45}$ & 0.47 / 0.16 / 0.38 \\
      & Einasto    & $-725$ & $8.87\times10^{-46}$ & 0.44 / 0.16 / 0.40 \\
    \multirow{2}{*}{endo, 1\,TeV}
      & SHM        & $+352$ & $1.37\times10^{-40}$ & 0.30 / 0.20 / 0.51 \\
      & Einasto    & $+324$ & $4.55\times10^{-39}$ & 0.33 / 0.28 / 0.39 \\
  \end{tabular}
  \end{ruledtabular}
\end{table}

Only $16$--$28\%$ of the counted events fall in W, and the rest land in S and H, where nothing was seen. This is why the curves lie below the LZ point. Raising $\sigma_p$ does not help, because it raises S and H by the same factor, and the empty regions then penalise the fit more than the extra events in W help it.

Xenon's second diffraction minimum is at $267\keV$, inside W. Every spectrum is therefore split into a lower lobe inside the ROI and an upper lobe above it, and only $0.03$--$0.05$ events are expected in $240$--$270\keV$ itself. The fits explain the event with the upper edge of the lower lobe plus the $32.5\keV$ uncertainty. For the exothermic SHM fit the kinematic peak is at $E_0 \simeq 325\keV$, inside the unobserved $270$-$350\keV$ gap.

Due to the lower cutoff speed, the Einasto spectrum ends near $490\keV$, as compared to $\sim590\keV$ for the SHM. If the SHM best-fit $(\delta,\sigma_p)$ are used unchanged in the Einasto halo, the exothermic prediction keeps its shape and simply scales with $\rho_\chi$ ($N_{\rm S}/N_{\rm W}/N_{\rm H} = 0.84/0.30/0.72$ at $0.567\GeV\,{\rm cm^{-3}}$). The endothermic prediction collapses to $\sim6\times10^{-7}$ events in total, because the SHM splitting puts $v_*$ only $6.6\,\rm km\,s^{-1}$ below the Einasto cutoff. The size of the suppression depends on the poorly known high-speed tail.

\begin{figure}[t]
  \centering
  \includegraphics[width=\columnwidth]{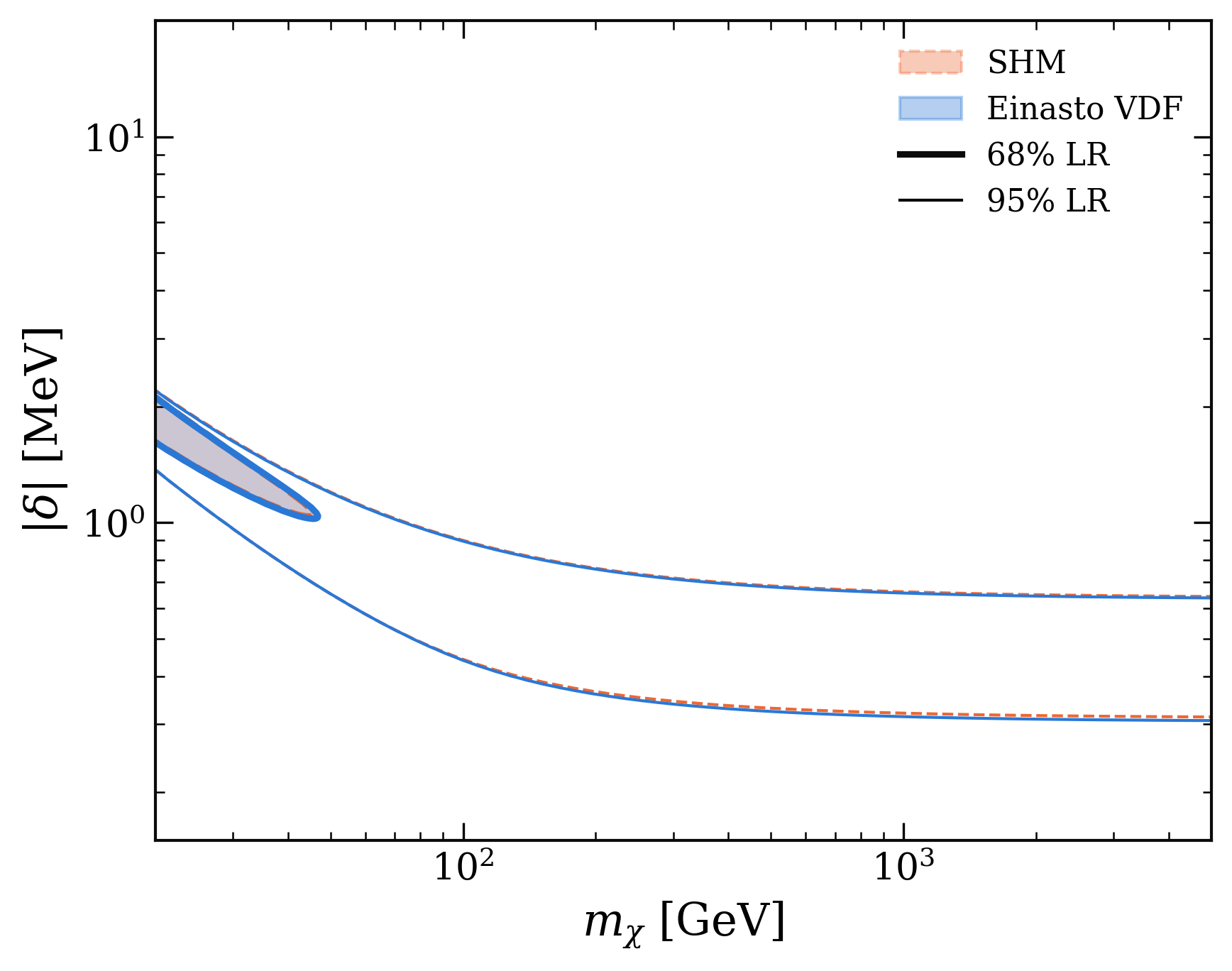}
  \caption{Allowed regions in the $m_\chi-\vert{}\delta\vert{}$ plane for exothermic scattering,
  with $\sigma_p$ profiled. These regions do not depend on $\rho_\chi$.}
  \label{fig:mdelta_exo}
\end{figure}

This is the same likelihood shown in the mass-splitting plane, with $\sigma_p$ profiled at every point (Figs.~\ref{fig:mdelta_exo} and~\ref{fig:regions_endo}).
Table~\ref{tab:notes-mdelta} lists the allowed intervals at representative masses.

\begin{table}[t]
  \caption{Allowed $\vert{}\delta\vert{}$ (keV) at fixed $m_\chi$, with $\sigma_p$ profiled, relative to each halo's best fit over the whole plane. The exothermic 68\% region exists only for $m_\chi \lesssim 45\GeV$, so above that only 95\% intervals are given.}
  \label{tab:notes-mdelta}
  \centering
  \scriptsize
  \begin{ruledtabular}
  \begin{tabular}{llcc}
    & $m_\chi$ & SHM & Einasto \\
    \hline
    exo (68\%)  & 30\,GeV  & $1236$--$1513$ & $1236$--$1513$ \\
    exo (95\%)  & 30\,GeV  & $963$--$1625$  & $963$--$1606$ \\
    exo (95\%)  & 100\,GeV & $444$--$896$   & $444$--$896$ \\
    exo (95\%)  & 300\,GeV & $347$--$716$   & $339$--$708$ \\
    exo (95\%)  & 1\,TeV   & $323$--$660$   & $316$--$652$ \\
    exo (95\%)  & 3\,TeV   & $318$--$641$   & $310$--$634$ \\
    \hline
    endo (68\%) & 500\,GeV & $302$--$327$ & $292$--$306$ \\
    endo (68\%) & 1\,TeV   & $312$--$363$ & $300$--$336$ \\
    endo (68\%) & 3\,TeV   & $318$--$375$ & $313$--$332$ \\
    endo (68\%) & 10\,TeV  & $321$--$378$ & $321$--$327$ \\
    endo (95\%) & 1\,TeV   & $202$--$363$ & $202$--$339$ \\
    endo (95\%) & 10\,TeV  & $198$--$392$ & $200$--$354$ \\
  \end{tabular}
  \end{ruledtabular}
\end{table}

For the exothermic case, the two halos give the same region: the best-fit $\vert{}\delta\vert{}$ agrees to within $2.4\%$ at every mass (median $1.2\%$), and the 95\% contours can hardly be told apart. The region is a diagonal band because the
event fixes the peak energy $E_0$, and the splitting needed to produce it, $\vert{}\delta\vert{} = E_0(1+m_N/m_\chi)$, grows towards low mass. At high mass $\vert{}\delta\vert{}\to E_0$ and the band flattens at $\vert{}\delta\vert{} \approx 300$--$650\keV$. The
best fit overall is at the lightest mass scanned, where the whole spectrum fits inside W ($F=1$). The 68\% region is therefore confined to $m_\chi \lesssim 42\GeV$ (SHM) and $46\GeV$ (Einasto), while the 95\% region has no
upper mass bound.

In the endothermic case, the allowed band is bounded above
by the kinematic edge, $\delta_{\rm edge}\propto\mu_{\chi N}v_{\rm cut}^2$ [Eq.~\eqref{eq:dedge}], and below by the requirement that the spectrum still reaches the event. A smaller cutoff speed ($768.5$ against $794.3\,\rm km\,s^{-1}$) therefore lowers the whole band. At $1\,{\rm TeV}$ the 68\% interval moves from $312$--$363\keV$ (SHM) to $300$--$336\keV$ (Einasto), and the best fit from
$+352$ to $+324\keV$. Over the mass range, the best-fit splitting is $6$--$31\%$ smaller with our VDF. The upper edge moves the most, as expected, since it is set directly by $v_{\rm cut}$. The band closes at low mass: the 68\% region starts at $m_\chi \approx 320\GeV$ (SHM) and $410\GeV$ (Einasto), the 95\% region at $200$ and
$300\GeV$. Below these masses, no splitting can place the spectrum at the event. Above a few TeV, the band flattens as $\mu_{\chi N}\to m_N$.

\section{Construction of the data-driven velocity distribution}
\label{app:vdf}

This appendix gives the details of the four steps summarised in the main text. The stellar catalogue is described in full in
Ref.~\cite{Mukherjee2026cat}. The rotation curve and the mass modelling are the subject of two companion papers in preparation
\cite{Mukherjee2026rc,Srivastav2026mass}, to which we refer for the
systematic tests that we can only summarise here.

\subsection{Stellar catalogue}
\label{app:catalogue}

Mapping the potential from the disc to the outer halo requires, for as many stars as possible, a distance and a full velocity. No single survey provides both over the whole Galaxy. Gaia measures positions and proper motions (the two velocity components on the sky) for over a billion stars, but its parallaxes lose precision beyond a few kpc, and its line-of-sight velocities
reach only bright stars. Ground-based spectroscopic surveys measure
line-of-sight velocities for fainter and more distant stars, each with its own footprint and calibration.

We therefore combined Gaia~DR3 \cite{Gaia2016,GaiaDR3} with
14 surveys, among them DESI \cite{DESI_DR1}, SDSS/SEGUE
\cite{SEGUE2009}, APOGEE \cite{APOGEE_DR17}, LAMOST \cite{LAMOST2012}, GALAH \cite{GALAH_DR4}, the Gaia-ESO survey \cite{GaiaESO2022} and RAVE \cite{RAVE_DR6}. Duplicate entries were merged star by star, and their velocities were combined using inverse-variance weights. Per-survey velocity zero points were fixed using stars observed by more than one survey. Parallaxes were corrected for the known Gaia zero-point offset \cite{Lindegren2021}, and beyond the parallax limit we used spectroscopic distances, standard candles (RR Lyrae and blue horizontal-branch stars, whose luminosities are known) and data-driven spectrophotometric distances \cite{Hogg2019}. Positions and velocities were transformed to Galactocentric coordinates with \textsc{Astropy} \cite{Astropy2013}, and uncertainties propagated by Monte Carlo sampling of each star's distance, proper motion and line-of-sight velocity.

The catalogue holds $32{,}552{,}876$ stars. The median fractional distance
uncertainty is $2.7\%$ within 15\,kpc of the Sun and $29\%$ beyond it, and the median fractional line-of-sight velocity uncertainty is $8.3\%$. It contains $517{,}123$ halo tracers, the largest homogeneous sample now available beyond the Gaia parallax limit \cite{Mukherjee2026cat}.

\subsection{Rotation curve}
\label{app:rc}

The circular speed $V_c(R)$ is defined by $V_c^2(R)=R\,\vert{}{\rm d}\Phi/{\rm d}R\vert{}$ for the Galactic potential $\Phi$, so it measures the enclosed mass. Real stars
do not move on circular orbits, so $V_c$ has to be recovered from the mean motions and the velocity dispersions of a population through the Jeans equations, which are the moments of the collisionless Boltzmann equation
\cite{BinneyTremaine2008}. We used three radial ranges.

Between 5 and 20\,kpc we used disc stars and the axisymmetric Jeans equation in cylindrical coordinates \cite{Eilers2019},
\begin{align}
  V_c^2 = \langle v_\phi^2\rangle
  &- \langle v_R^2\rangle\left(1+\frac{\partial\ln\nu}{\partial\ln R}
   +\frac{\partial\ln\langle v_R^2\rangle}{\partial\ln R}\right) \nonumber\\
  &- R\,\frac{\partial\langle v_Rv_z\rangle}{\partial z},
\end{align}
where $\nu$ is the number density of the tracer stars and $v_R,v_\phi,v_z$ are the velocity components in cylindrical coordinates. The second term is the asymmetric drift (a population with random motions rotates more slowly than the
circular speed). Between 20 and 250\,kpc too few stars have precise proper motions, so we used line-of-sight velocities only. For each tracer class we measured the line-of-sight velocity dispersion in the Galactic rest frame, $\sigma_{\rm GSR}(r)$, in radial bins, and the slope $\alpha_n$ of the tracer density, $\nu\propto r^{-\alpha_n}$. The spherical Jeans equation then gives
\cite{Xue2008,Bhattacharjee2014}
\begin{equation}
  V_c^2(r) = \sigma_r^2\left(\alpha_n+2\gamma_\sigma-2\beta\right),
\end{equation}
with $\gamma_\sigma=-{\rm d}\ln\sigma_r/{\rm d}\ln r$ and $\beta$ the velocity anisotropy, which measures how much the random motions favour radial over
tangential directions. Our baseline curve takes $\beta=0$, for which the
line-of-sight and radial dispersions coincide. Inside 5\,kpc the bar makes
stellar kinematics hard to interpret, and we used published gas terminal
velocities instead \cite{Bhattacharjee2014,Sofue2020}.

All velocities depend on the adopted Local Standard of Rest, i.e., on
$(R_0,V_0)$. Curves quoted at some other $(\tilde R_0,\tilde V_0)$ were
converted with
\begin{equation}
  V_c(R)\to\frac{\tilde R_0}{R_0}
  \left[V_c(R)-\frac{R}{\tilde R_0}(\tilde V_0-V_0)\right],
\end{equation}
which leaves the tangential velocity in a non-rotating frame unchanged. We
repeated everything for $(R_0,V_0)=(8.0,200)$, $(8.1,229)$, $(8.3,244)$ and $(8.5,220)$ in kpc and $\kms$ \cite{KerrLyndenBell1986,GRAVITY2018,Eilers2019}. The results in this paper use $(8.1\,{\rm kpc},229\kms)$. An extra systematic uncertainty of $8\kms$ was added in quadrature to every point, to allow for
zero-point offsets between tracers measured in different ways.

\subsection{Mass model}
\label{app:massmodel}

We fitted the curve with a spherical DM halo plus baryons. Four halo profiles were tried: NFW \cite{Navarro1997}, Burkert \cite{Burkert1995}, a cored
pseudo-isothermal sphere, and Einasto \cite{Einasto1965},
\begin{equation}
  \rho_\chi(r)=\rho_s\exp\left\{-\frac{2}{\alpha}
  \left[\left(\frac{r}{r_s}\right)^{\alpha}-1\right]\right\}.
\end{equation}
The baryons were described by nine configurations built from one or two
exponential stellar discs \cite{Freeman1970}, a gas disc, and a Plummer \cite{Plummer1911} or Hernquist \cite{Hernquist1990} bulge. Masses and scale lengths carried Gaussian priors centered on standard MW values ($4\times10^{10}M_\odot$ for the stellar disc, $1.3\times10^{10}M_\odot$ for
the gas, $9\times10^{9}M_\odot$ for the bulge), and halo concentrations a prior from the mass-concentration relation of $\Lambda$CDM simulations
\cite{Dutton2014}.

Each of the $4\times9=36$ combinations was fitted at each of the four $(R_0,V_0)$ choices with the ensemble sampler \textsc{emcee}
\cite{ForemanMackey2013} and a Gaussian likelihood. Chains were run until the Gelman-Rubin statistic, which compares the scatter within each chain to the scatter between chains, satisfied $\hat R-1<0.01$ for every parameter \cite{GelmanRubin1992}. Models were ranked by the widely applicable information criterion \cite{Watanabe2010}, which rewards goodness of fit and penalises
unnecessary freedom. Einasto halo fits best at every $(R_0,V_0)$. The adopted fit gives $M_{200}=(4.14\pm0.14)\times10^{11}M_\odot$, $r_s=10.2\pm0.4$\,kpc,
$\alpha=0.494^{+0.005}_{-0.010}$ and $r_{200}\simeq151$\,kpc. Across the four LSR conventions, $M_{200}$ stays within $(4.1$--$4.4)\times10^{11}M_\odot$ and
$\rho_\chi(R_0)$ within $0.54$--$0.60\GeV\,{\rm cm^{-3}}$
\cite{Srivastav2026mass}. This virial mass sits at the low end of the
literature, which reflects the fairly low circular speeds we measure beyond
50\,kpc and the fact that an Einasto halo, unlike NFW, has a finite total mass.

\subsection{Eddington inversion}
\label{app:eddington}

From the fitted model we build the total circular speed $V_c^2=V_\chi^2+V_{\rm disc}^2+V_{\rm gas}^2+V_{\rm bulge}^2$, treat it as
spherical, and integrate it to get $\Psi(r)$. The Eddington integral in
Eq.~\eqref{eq:eddington} needs a halo with a finite total mass and a smooth
outer edge. A sharp cut would produce a spike in
${\rm d}^2\rho_\chi/{\rm d}\Psi^2$. We therefore keep the fitted density
unchanged out to $r_{200}$ and multiply it beyond that by
$\exp\{-[(r-r_{200})/r_{200}]^2\}$.

Numerically, $\rho_\chi$ and $\Psi$ are tabulated on 2400 logarithmically
spaced radii, and $\ln\rho_\chi$ is splined against $\Psi$ and differentiated twice, which keeps the derivatives well behaved across many orders of magnitude in density. The integrand of Eq.~\eqref{eq:eddington} diverges at $\Psi=\mathcal{E}$. The substitution $\Psi=\mathcal{E}-q^2$ removes the divergence \cite{Widrow2000}. We checked the inversion against the two
cases with closed-form answers, it reproduces the Maxwellian of a singular isothermal sphere to $10^{-5}$ and the exact $f(\mathcal{E})$ of a Plummer
sphere to $10^{-7}$ \cite{BinneyTremaine2008}.

The boost to the laboratory frame uses the isotropy of $f(v)$ in the Galactic frame, which reduces Eq.~\eqref{eq:eta} to a one-dimensional integral. Using the
truncated Maxwellian of the SHM, our implementation reproduces the standard
analytic result for $\eta$ to better than $0.2\%$ at all $\vmin$.

\begin{figure}[t]
  \centering
  \includegraphics[width=\columnwidth]{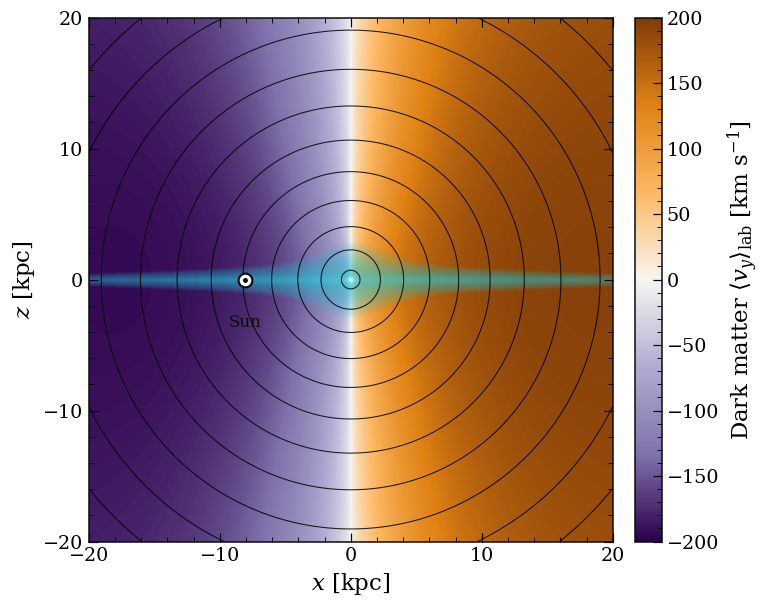}
  \caption{Dark matter mean lab-frame velocity component $\langle v_y\rangle_{\rm lab}$ (colour scale) in the Galactocentric $x$--$z$ plane. The black circles are centred on the Galactic centre, the position of the Sun ($R_0=8.1$\,kpc) is marked, and the teal band around $z=0$ shows the flattened baryonic disc and bulge.}
  \label{fig:halo_map}
\end{figure}

Figure~\ref{fig:halo_map} gives a picture of the geometry behind this construction. It shows the DM halo in the Galactocentric $x$--$z$ plane through the Sun (at $x=-R_0$, $z=0$), coloured by the mean lab-frame velocity component $\langle v_y\rangle_{\rm lab}$. The halo is spherical, as assumed above, so the black circles centred on the Galactic centre are the natural surfaces on which its properties are constant, and the local distribution $4\pi v^2f(v)$ of Fig.~\ref{fig:vdf} is the one evaluated on the circle through the Sun. The map vanishes along $x=0$, changes sign across it, and reaches $\pm200\kms$ at $|x|\simeq20$\,kpc. The baryonic disc, which is much smaller than the halo and enters the potential only through $V_c(R)$, is confined to the thin band around $z=0$. This is why treating the disc as a spherical contribution to the potential (see Sec.~\ref{app:robust}) is a good approximation for the halo as a whole, though not necessarily near the disc plane.

\begin{figure}[t]
  \centering
  \includegraphics[width=\columnwidth]{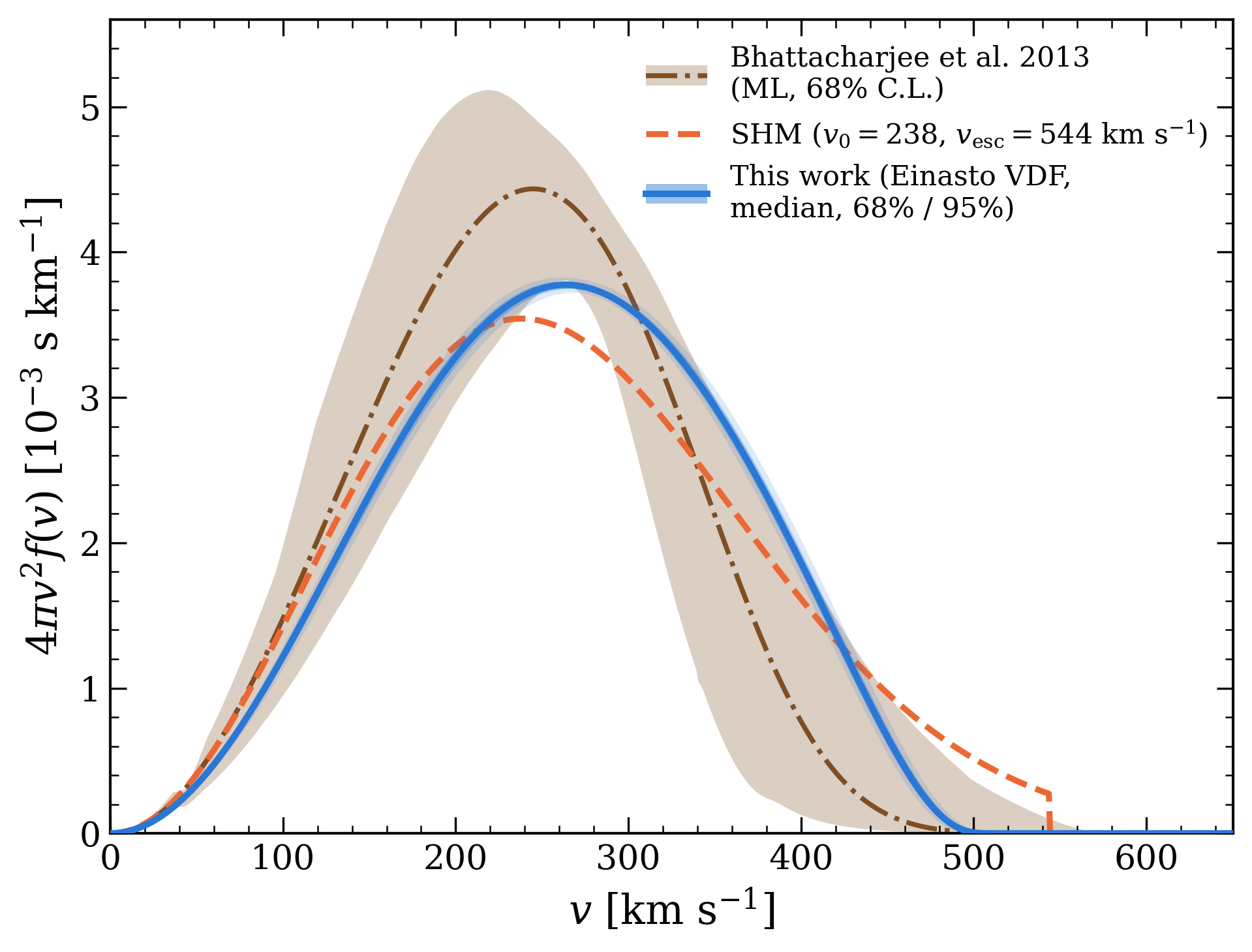}
  \caption{Local DM speed distribution for the Einasto halo (blue; median with 68\% and 95\% posterior bands) compared with the earlier rotation-curve-derived result of Bhattacharjee et al.~\cite{Bhattacharjee2013}
  (brown; most-likely fit with 68\% C.L. band) and the Standard Halo Model (orange dashed). The uncertainty band on $\vesc$ is narrower by a factor of $\sim38$, reflecting the larger stellar catalogue, extended rotation curve
  and systematic mass-model comparison of this work relative to the single MCMC-fitted NFW halo of Ref.~\cite{Bhattacharjee2013}.}
  \label{fig:vdf_bhatt}
\end{figure}

\subsection{Comparison with the earlier rotation-curve-derived VDF}
\label{app:vdf-compare}

The same four-step logic used here (catalogue, rotation curve, mass model, Eddington inversion) was applied earlier by Bhattacharjee et al.~\cite{Bhattacharjee2013}, who fitted a single MCMC-sampled NFW halo to an older, shorter rotation-curve compilation and propagated the resulting 68\% credible interval on the halo parameters through the same Eddington integral to obtain a most-likely $4\pi v^2f(v)$ curve with an uncertainty band. This is, to our knowledge, the closest earlier analogue to the present construction, and it provides a direct way to see what the larger catalogue, the extended rotation curve and the systematic mass-model comparison of Secs.~\ref{app:catalogue}--\ref{app:massmodel} actually buy. Figure~\ref{fig:vdf_bhatt} overlays their most-likely curve and 68\% C.L. band, the SHM, and the data-driven distribution of this work. The 68\% C.L. width of $\vesc$ shrinks from $^{+121}_{-98}\kms$ in Ref.~\cite{Bhattacharjee2013} to $\pm2.9\kms$ here, a factor of $\sim38$, while the two most-likely curves agree on the bulk of the distribution to within their respective uncertainties.

\subsection{Robustness}
\label{app:robust}
The construction assumes that the halo is spherical, isotropic and in
equilibrium, and treats the flattened disc as a spherical contribution to the
potential. Tests against hydrodynamical simulations show that Eddington
inversion recovers the bulk of the local speed distribution, while the
agreement in the tail is model-dependent and the simulations themselves do not
agree with one another there \cite{Lacroix2020,Evans2019}. The same
spread is visible between rotation-curve-based distributions built with different mass models. \cite{Bhattacharjee2013}.

Therefore, the overall scale of the
tail, set by $\vesc$, is robust, as the rotation curve constrains the potential
out to the virial radius, and changing the outer truncation of the halo by a
factor of eight moves $\vesc$ by less than $10^{-3}\%$. The \emph{shape} of the
distribution within the last few tens of $\kms$ below $\vesc$ is not robust, since
the same change alters the rate from particles within $20\kms$ of $v_{\rm cut}$
by more than $5\%$, and by up to an order of magnitude for particles within a few $\kms$ of it. The sharp SHM cutoff is an
assumption of exactly the same kind, and neither model can exclude a small
population of unbound stars-like debris above $\vesc$. For this reason we
discard, for both halo models, any fit that draws its signal from DM within
$20\kms$ of $v_{\rm cut}$ (the edge rule in Table~\ref{tab:notes-inputs}).
\end{document}